\documentclass[iop,numberedappendix,revtex4]{emulateapj}
\usepackage{natbib}
\usepackage{apjfonts}
\usepackage{aas_macros}
\def\baselinestretch{1.0}
\usepackage{amsmath}
\usepackage{color}

\usepackage{hyperref}

\usepackage{pdflscape}

\def\sigv{\langle \sigma v \rangle}
\def\Mchi{M_{\chi}}
\def\nhat{\hat{\mathbf{n}}}
\def\MMW{M_{\rm MW}}
\def\Prob{{\rm P}}
\def\CDF{{\rm CDF}}
\def\degg{^\circ}
\def\MMW{M_{\mathrm{MW}}}
\def\vmax{v_{\rm max}}
\def\rmax{r_{\rm max}}
\def\Jdecay{J_{\mathrm{decay}}}

\def\cv{c_{{\mathrm{v}}}}
\def\thetamax{\theta_{\mathrm{max}}}
\def\J{J}
\def\deg{{\mathrm{deg}}}

\begin{document}

\title{Dwarf galaxy annihilation and decay emission profiles for dark matter experiments }
\shorttitle{Dwarf galaxy profiles for dark matter experiments}
\shortauthors{Geringer-Sameth, Koushiappas \& Walker}

\author{Alex Geringer-Sameth\altaffilmark{1}}
\author{Savvas M. Koushiappas\altaffilmark{2}}
\author{Matthew Walker\altaffilmark{3}}

\altaffiltext{1}{Department of Physics, Brown University, Providence, RI 02912 and McWilliams Center for Cosmology, Department of Physics, Carnegie Mellon University, Pittsburgh, PA 15213; alexgs@cmu.edu}
\altaffiltext{2}{Department of Physics, Brown University, Providence, RI 02912; koushiappas@brown.edu}
\altaffiltext{3}{McWilliams Center for Cosmology, Department of Physics, Carnegie Mellon University, Pittsburgh, PA 15213; mgwalker@andrew.cmu.edu}

\begin{abstract}
Gamma-ray searches for dark matter annihilation and decay in dwarf galaxies rely on an understanding of the dark matter density profiles of these systems. Conversely, uncertainties in these density profiles propagate into the derived particle physics limits as systematic errors. In this paper we quantify the expected dark matter signal from 20 Milky Way dwarfs using a uniform analysis of the most recent stellar-kinematic data available.  Assuming that the observed stellar populations are equilibrium tracers of spherically-symmetric gravitational potentials that are dominated by dark matter, we find that current stellar-kinematic data can predict the amplitudes of annihilation signals to within a factor of a few for the ultra-faint dwarfs of greatest interest. On the other hand, the expected signal from several classical dwarfs (with high-quality observations of large numbers of member stars) can be localized to the $\sim20\%$ level. These results are important for designing maximally sensitive searches in current and future experiments using space and ground-based instruments.
\end{abstract}

\keywords{dark matter --- galaxies: dwarf --- galaxies: fundamental parameters --- galaxies: kinematics and dynamics  }

\section{Introduction}

The search for cosmological dark matter annihilation or decay is a major effort in contemporary astrophysics. Educing the dark matter particle physics from observations requires a detailed understanding of the dark matter distribution in the systems under study. A productive avenue of approach has been to search for gamma-rays generated by dark matter annihilation in Milky Way dwarf spheroidal galaxies \citep[e.g.][]{2010JCAP...01..031S,2010PhRvD..82l3503E,2011JCAP...06..035A,2011PhRvL.107x1303G,2011PhRvL.107x1302A,2012PhRvD..86b1302G,2012PhRvD..85f2001A,2014PhRvD..89d2001A,2014arXiv1410.2242G}.  Such systems are nearby, dark matter-dominated, and contain no conventional sources of astrophysical backgrounds (e.g. cosmic ray generation and propagation through interstellar gas). Many such dwarf galaxies have been discovered in recent years \citep{willman05a,zucker06a,zucker06b,walsh07,belokurov07,belokurov08,belokurov09,belokurov10} with the prospect of more discoveries from ongoing and future sky surveys like Pan-Starrs \citep{2002SPIE.4836..154K}, the Vista Hemisphere Survey \citep{2013ApJS..209...22A,2014ApJS..212...16A}, the Dark Energy Survey \citep{2005IJMPA..20.3121F}, and eventually the Large Synoptic Survey Telescope \citep{2003NuPhS.124...21T}.

Previous studies of dwarf galaxies have begun to constrain the physical properties of dark matter \citep{2011PhRvL.107x1303G,2011PhRvL.107x1302A,2014PhRvD..89d2001A,2014arXiv1410.2242G}. The lack of any significant gamma-ray excess lead to the exclusion of generic dark matter candidates with annihilation cross sections on the order of the benchmark value for a thermal relic ($\sim 3 \times 10^{-26} {\mathrm{cm^3 s^{-1}}}$) and with masses less than a few tens of GeV. Despite current non-detections, dwarf galaxies---and their lack of astrophysical contaminating sources---offer the cleanest possible signature of dark matter annihilation or decay compared with other targets. This is especially interesting in the context of recent claims of a Galactic center gamma-ray excess and associated dark matter interpretation~\citep[e.g.][]{2011PhLB..697..412H,2011PhLB..705..165B,2012PhRvD..86h3511A,2013PhRvD..87l9902A,2014PhRvD..90b3526A,2014arXiv1402.6703D}.  Observations of dwarf galaxies have the potential to either confirm or rule out such an interpretation. 

The dark matter distribution within a target system is a necessary ingredient for placing constraints on any particle theory that predicts dark matter annihilation or decay. Knowledge of the relative signal strengths amongst different targets as well as the spatial distribution of the emission is required for designing maximally sensitive searches in current and future experiments. The overall emission rate from annihilation is described by the ``$J$ value'', the integral along the line-of-sight and over an aperture of the square of the dark matter density. The amplitude of $J$ helps to identify which dwarfs are the most promising for searches (i.e. are the ``brightest'').  

Different groups have devised various methods for estimating dark matter distributions (and uncertainties) using observations of line-of-sight velocities of dwarf galaxy member stars.  Some authors use the kinematic data to fit for the mass and/or concentration of dark matter density profiles that are assumed to follow an analytic form typically used to describe low-mass ``subhalos'' (virial mass $\sim 10^{9-10}M_{\odot}$) that form around  Milky-Way-like galaxies in dissipationless cosmological simulations based on cold dark matter~\citep[e.g.][]{2007PhRvD..75h3526S,2009JCAP...06..014M,2013arXiv1309.2641M}.  Some studies make an explicit assumption of a cored profile~\citep{2012PhRvD..86b3528C,2012MNRAS.420.2034S}, while others take a more agnostic approach, fitting relatively flexible density profiles that are not restricted to the form used to describe simulated halos \citep[e.g.][]{2011MNRAS.418.1526C}. 

In addition, different groups use different techniques for propagating the uncertainties in those dark matter distributions when incorporating gamma-ray non-detections to derive limits on the annihilation cross section as a function of particle mass. For example, in their joint analysis of stellar-kinematic and gamma-ray data for several dwarfs, \citet{2011PhRvL.107x1302A} take the uncertainty in $J$ to be described by a log-normal distribution that is subsequently folded into a likelihood for a given particle physics model.  Working with similar data, \citet{2011PhRvL.107x1303G} separated that same log-normal distribution for $J$ from statistical uncertainties in the gamma-ray data, allowing this systematic $J$ uncertainty to be compared with the derived cross section limits. The results shown by \citet{2011PhRvL.107x1303G} are alarming: the uncertainty in $J$ can affect limits on the cross section by a factor of 10; nevertheless, the conservative edge of the 95\% systematic band still rules out light dark matter with the benchmark thermal cross section (see Fig. 2 in \cite{2011PhRvL.107x1303G}).

Building on these previous works, we have devised a novel statistical approach that operates on stellar-kinematic and gamma-ray data in order to maximize sensitivity to annihilation/decay signals.  We present results in a pair of papers.  Here (Paper I), we use published stellar-kinematic data to estimate the dark matter density profiles of 20 Milky Way dwarfs.  These systems lie at heliocentric distances $23 \la D/\mathrm{kpc}\la  250$ and span a range in luminosity $3\la \log_{10}[L/L_{\odot}]\la 7$ \citep[and references therein]{2012AJ....144....4M}.  The primary goal of this work is to quantify uncertainties in the dark matter density profiles---and hence in the $J$ values---that reflect statistical errors due to finite data as well as uncertainties regarding the shapes of dark matter density profiles.  We adopt the relatively agnostic modeling approach of \citet{2011MNRAS.418.1526C} and extend it to the least luminous of the Milky Way's dwarf satellites, where statistical uncertainties can become dominant over systematics; for a thorough examination of the behaviors of random and systematic errors that are inherent to this analysis, we refer the reader to the recent study by \citet{2015MNRAS.446.3002B}.  In Paper II, we present a new analysis of gamma-ray data from the Fermi \textit{Large Area Telescope} (LAT) that, in combination with the estimates of $J$ values presented here, places new limits on dark matter particle interactions.

This paper is structured as follows. In Sec.~\ref{sec:darkmattersignal} we introduce the quantities we seek to constrain by briefly reviewing the physics of dark matter annihilation and decay. Sec.~\ref{sec:estimates} describes the parameterization of the density profile and the equations relating this profile to astronomical observables. Sec.~\ref{sec:observations} introduces the observational datasets used in this analysis and Sec.~\ref{sec:fitting} describes the fitting procedure. We discuss some physical considerations and the importance of truncating the dark matter potentials in Sec.~\ref{sec:constraints}. The results of the analysis are presented in Sec.~\ref{sec:results}.  We discuss their relevance in the context of previous work in Sec.~\ref{sec:others} and conclude in Sec.~\ref{sec:conclusion}. Appendix~\ref{sec:appendix} presents constraints on integrated emission profiles and containment fractions for each dwarf, which are also provided in machine-readable form.

\section{Dark matter signal}
\label{sec:darkmattersignal}

The distribution of dark matter within a system determines the flux of photons and other products of dark matter annihilation and decay. The annihilation rate per volume per time is given by 
\begin{equation} 
R_{\mathrm {ann}} = \frac{1}{2} n^2 \sigv,
\label{eq:gammaann} 
\end{equation}
where $n$ is the number density of dark matter particles and $\sigv$ is the velocity-averaged annihilation cross section~\citep[e.g.][]{1996PhR...267..195J, 1991NuPhB.360..145G}. (Note that $n$ represents the total number density of dark matter irrespective of particle vs. antiparticle; an additional factor of $\frac{1}{2}$ appears if particles and antiparticles each constitute half the total abundance.)
As dynamical observations offer a handle on the mass density of dark matter but not its number density, it is convenient to write $n = \rho/\Mchi$, where $\rho$ is the mass density and $\Mchi$ is the mass of a single dark matter particle.

If dark matter decays, the decay rate per volume is given by 
\begin{equation} 
R{_{\mathrm{dec}}} = n \Gamma,
\end{equation}
where $\Gamma$ is the decay rate of an individual dark matter particle. 

Each annihilation (or decay) gives rise to photons described by the spectrum $dN_\gamma (E)/dE$: the number of photons per energy produced per annihilation (or decay). Photons produced in annihilation or decay travel along straight lines and so the expected flux of photons is determined by dark matter annihilation (or decay) taking place along the entire line of sight in a certain direction. The number of photons per solid angle per energy, area, and time coming from sky-direction $\nhat$, in the case of annihilation, is given by
\begin{equation}
\frac{dF(\nhat, E)}{d\Omega dE} = \frac{\sigv}{8\pi \Mchi^2} \frac{dN_\gamma (E)}{dE}
                                              \int_{\ell=0}^\infty d\ell \left[ \rho(\ell \nhat) \right]^2,
\label{eqn:dF_annihilation}
\end{equation}
while for decay it is given by
\begin{equation}
\frac{dF(\nhat, E)}{d\Omega dE} = \frac{\Gamma}{4\pi \Mchi} \frac{dN_\gamma (E)}{dE}
                                              \int_{\ell=0}^\infty d\ell \, \rho(\ell \nhat).
\label{eqn:dF_decay}
\end{equation}
Here, $F$ has the units of photons per area per time, $\ell$ is the line of sight distance from Earth and $\rho(\ell \nhat)$ is the dark matter mass density at location $\ell \nhat$. The integrals over the line of sight should be thought of as including the units of [solid angle]$^{-1}$. For the case of dark matter annihilation, we define the $J$-profile to be
\begin{equation}
\frac{dJ(\nhat)}{d\Omega} =  \int_{\ell=0}^\infty d\ell \left[ \rho(\ell \nhat) \right]^2.
\label{eqn:Jdef}
\end{equation}
The corresponding quantity in the case of dark matter decay is the projected mass density along the line of sight:
\begin{equation}
\frac{d\Jdecay(\nhat)}{d\Omega} =  \int_{\ell=0}^\infty d\ell \, \rho(\ell \nhat).
\label{eqn:Jdecaydef}
\end{equation}

The terms before the integrals in Eqs.~\eqref{eqn:dF_annihilation} and~\eqref{eqn:dF_decay} describe the microscopic physics of dark matter while the $J$-profile reflects its distribution on large scales.  The goal of indirect detection is to use knowledge of $dJ/d\Omega$ (or $d\Jdecay/d\Omega$) along with observations of photons ($dF/d\Omega dE$) to learn something about dark matter particle physics $\{\Mchi, \sigv, \Gamma\}$. In the following we will present results for both annihilation and decay.

We consider spherically symmetric density profiles and so $dJ/d\Omega$ is a function only of $\theta$, the angular separation between the line of sight $\nhat$ and the direction towards the center of the dwarf galaxy. The integration in Eqs.~\eqref{eqn:Jdef} and~\eqref{eqn:Jdecaydef} along the line of sight is carried out numerically for every value of $\theta$. 
We let the variable $x$ denote the distance, along the line of sight, from the point where the line of sight makes its closest approach to the center of the dwarf. That is, the line of sight corresponding to an angular separation $\theta$ has an impact parameter $b = D \sin(\theta)$, where $D$ is the distance from Earth to the center of the dwarf. The limits of integration in Eqs.~\eqref{eqn:Jdef} and~\eqref{eqn:Jdecaydef} are $x=-\infty$ and $x=+\infty$, and $dx=d\ell$. The dark matter density is a function of $r=\sqrt{b^2 + x^2}$, the distance from the center of the dwarf, so that $\rho(\ell \nhat)$ is given by $\rho(\sqrt{b^2+x^2})$.

\section{Reconstructing the dark matter potential with stellar kinematics}
\label{sec:estimates}

\subsection{Dark matter density}

In order to accurately quantify uncertainties in the spatial distribution of dark matter it is necessary to use a suitably flexible functional form for the density profile \citep{2015MNRAS.446.3002B}.  Following \citet{2011MNRAS.418.1526C}, we adopt the functional form introduced by \citet{1996MNRAS.278..488Z} to generalize the \citet{1990ApJ...356..359H} profile.  In this spherically symmetric model, the density of dark matter at halo-centric radius $r$ is
\begin{equation}
\rho(r) = \frac{\rho_s}{\left( r / r_s \right)^\gamma   \left[1 + \left( r / r_s \right)^\alpha \right] ^{ ( \beta -\gamma) / \alpha }}.
\label{eq:rho}
\end{equation}
This five-parameter profile, normalized by the scale density $\rho_s$, describes a split power law with inner logarithmic slope $d\log\rho/d\log r|_{r\ll r_s}=-\gamma$ and outer logarithmic slope $d\log\rho/d\log r|_{r\gg r_s}=-\beta$.  The transition happens near the scale radius $r_s$, with $\alpha$ specifying its sharpness.  For $(\alpha,\beta,\gamma)=(1,3,1)$ one recovers the two-parameter NFW profile that characterizes cold dark matter (CDM) halos formed in dissipationless numerical simulations \citep{1997ApJ...490..493N}.  However, the profile can also describe halos with even steeper central ``cusps'' ($\gamma>1$), or halos with ``cores'' of uniform central density ($\gamma \sim 0$), as are usually inferred from observations of real galaxies \citep[and references therein; \citealt{walker11,2009MNRAS.397.1169D}]{2010AdAst2010E...5D}.  This flexibility lets us explore a wide range of physically plausible dark matter profiles.

\subsection{Estimation of dark matter profile parameters}

From Eq.~\eqref{eqn:dF_annihilation}, the flux of annihilation by-products depends on the density of dark matter particles within the source, and thus on the source's gravitational potential.  For collisionless stellar systems like dwarf galaxies, the gravitational potential is related fundamentally to the phase-space density of stars $f(\bf{r},\bf{u})$, defined such that $f({\bf{r}},{\bf{u}})\, d^3{\bf{r}}\, d^3{\bf{u}}$ gives the expected number of stars lying within the phase-space volume $d^3{\bf{r}}\, d^3{\bf{u}}$ centered on $(\bf{r}, \bf{u})$.  However, dwarf galaxies are sufficiently far away that current instrumentation resolves only the projection of their internal phase-space distributions, effectively providing information in just three dimensions: position as projected onto the plane perpendicular to the line of sight, and velocity along the line of sight (from Doppler redshift).  Given these limitations, it is common to infer the gravitational potential $\Phi$ by considering its relation to moments of the phase-space distribution: the stellar density profile, 
\begin{equation}
\nu(r)\equiv\int f({\bf{r}},{\bf{u}}) \,\, d^3{\bf{u}},
\end{equation}
and the stellar velocity dispersion profile, 
\begin{eqnarray}
\overline{u^2}(r) &=& \overline{u_r^2}(r)+\overline{u_{\theta}^2}(r) + \overline{u_{\phi}^2}(r) \\
                             &=& \frac{1}{\nu(r)}\int u^2 f({\bf{r}},{\bf{u}}) \,\, d^3{\bf{u}}.  
\end{eqnarray}
Assuming dynamic equilibrium and spherical symmetry, these quantities are related according to the spherical Jeans equation \citep{2008gady.book.....B},
\begin{equation}
  \frac{1}{\nu(r)}\frac{d}{dr}[\nu(r) \overline{u_r^2}(r)]+2\frac{\beta_a(r)\overline{u_r^2}(r)}{r}=-\frac{d\Phi}{dr}=-\frac{GM(r)}{r^2},
  \label{eq:jeans}
\end{equation}
where 
\begin{equation}
\beta_a(r)\equiv 1-\frac{\overline{2u_{\theta}^2}(r)}{\overline{u_r^2}(r)}
\label{eq:beta_a}
\end{equation}
characterizes the orbital anisotropy and the enclosed mass profile 
\begin{equation} 
M(r)=4\pi\int_{0}^{r} s^2\rho(s)ds
\end{equation} 
includes contributions from the dark matter halo.  

Equation \ref{eq:jeans} has the general solution \citep{vandermarel94,mamon05}
\begin{equation}
  \nu(r)\overline{u^2_r}(r)=\frac{1}{f(r)}\displaystyle\int_{r}^{\infty}f(s) \, \nu(s) \,\frac{GM(s)}{s^2} \,ds,
  \label{eq:jeans2}
\end{equation}
where 
\begin{equation} 
f(r)=2 \, f(r_1) \, \exp\left[ \int_{r_1}^{r}\beta_a(s)s^{-1} \, \, ds \right].
\end{equation}
Projecting along the line of sight, the mass profile relates to observable profiles, the projected stellar density $\Sigma(R)$, and line-of-sight velocity dispersion $\sigma(R)$, according to \citep{2008gady.book.....B}
\begin{equation}
  \sigma^2(R) \, \Sigma(R)=2 \, \displaystyle \int_{R}^{\infty}\biggl (1-\beta_a(r)\frac{R^2}{r^2}\biggr ) \frac{\nu(r) \, \, \overline{u_r^2}(r) \, \, r}{\sqrt{r^2-R^2}} \, \, dr.
  \label{eq:jeansproject}
\end{equation}

We use Eq.~\eqref{eq:jeansproject} to fit models for $\rho(r)$ and $\beta_a(r)$ to observed velocity dispersion and surface brightness profiles under the following assumptions:

\begin{itemize}
\item dynamic equilibrium and spherical symmetry, both implicit in the use of Eq.~\eqref{eq:jeans};
\item the stars are distributed according to a \citet{plummer11} profile, 
\begin{equation}
\nu(r)=\frac{3L}{4\pi R_e^3}\frac{1}{(1+R^2/R_e^2)^{5/2}},
\end{equation}
implying surface brightness profiles of the form 
\begin{equation}
\Sigma(R)=\frac{L}{\pi R_e^2}\frac{1}{(1+R^2/R_e^2)^2},
\end{equation}
where $L$ is the total luminosity and $R_e$ is the projected halflight radius;  
\item the stars contribute negligibly to the gravitational potential, such that $R_e$ is the only meaningful parameter in $\nu(r)$ and $\Sigma(R)$;
\item $\beta_a = {\rm constant}$;
\item the distribution of stellar velocities is not significantly influenced by the presence of binary stars.
\end{itemize}
Real galaxies violate all of these assumptions at some level and it is important to consider that the error distributions that we derive for $J$ values will not include the resulting systematic errors.  For the present work, we are concerned primarily with quantifying statistical uncertainties that arise from finite sizes of stellar-kinematic samples.  For a thorough study of systematic errors that can arise due to different stellar density profiles, non-spherical symmetry, and more complicated behaviors of the velocity anisotropy, we refer the reader to the recent study by \citet{2015MNRAS.446.3002B}.

\section{Observations}
\label{sec:observations}

\subsection{Classical dwarfs}
\label{subsec:classical}

For the Milky Way's eight most luminous ``classical'' dwarf galaxies, we adopt projected halflight radii listed in Table 1 of \citet[][original source is \citealt{ih95}]{walker09a}.  We use the stellar-kinematic data published by \citet{2008ApJ...675..201M} for Leo I, and by \citet{walker09a} for Carina, Fornax, Sculptor and Sextans.  

For Draco, Leo II, and Ursa Minor we use stellar-kinematic data acquired with the Hectochelle spectrograph at the 6.5-m MMT.  These data have previously been analyzed by \citet{2009ApJ...704.1274W} and \citet{2011MNRAS.418.1526C}, and will soon be made public (Walker, Olszewski \& Mateo, in preparation).

\begin{figure*}
\begin{center}
  \includegraphics[scale=1.]{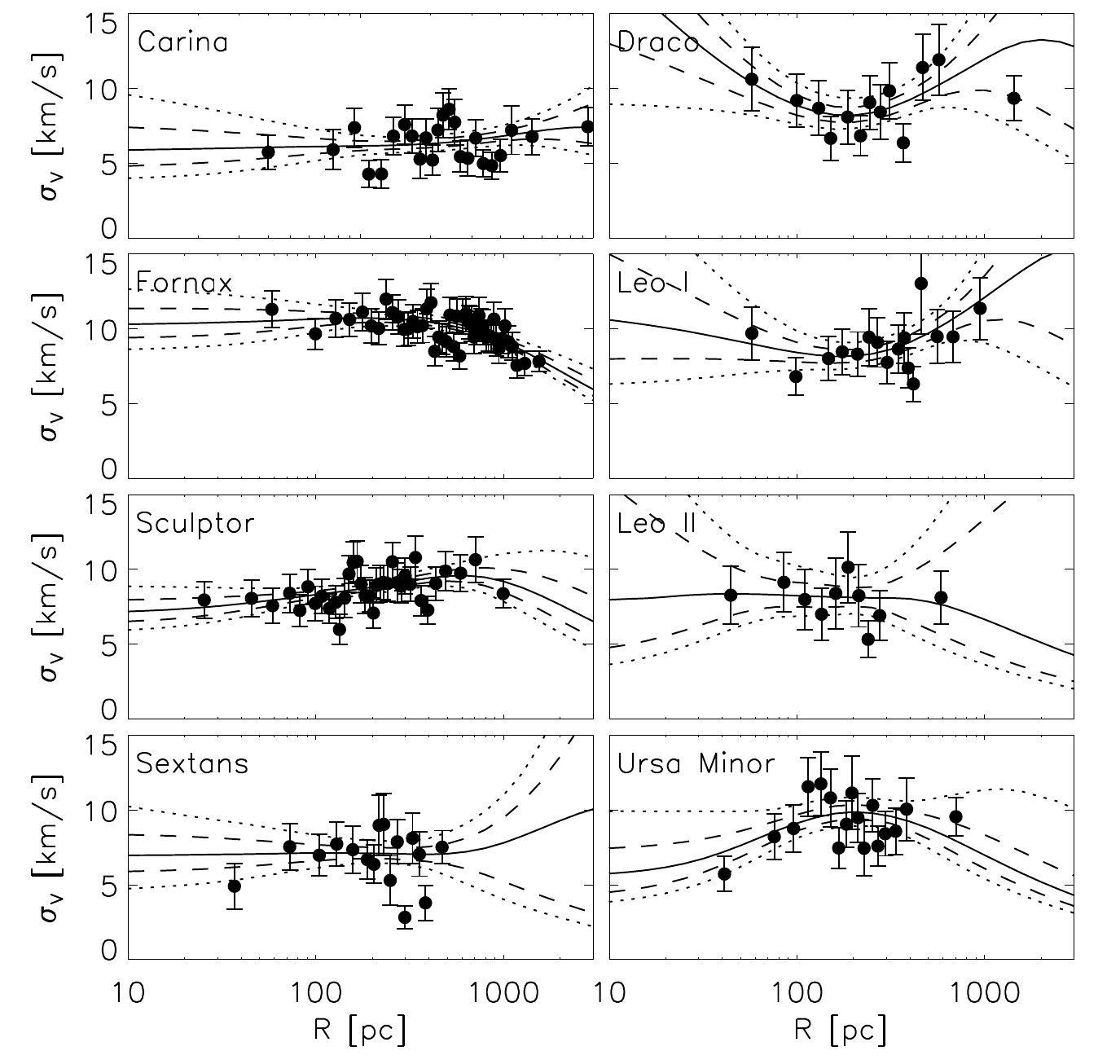}
  \caption{Line-of-sight stellar velocity dispersion profiles observed for the Milky Way's eight classical dwarf spheroidal satellites, adopted from \citet{2009ApJ...704.1274W}.  Solid curves indicate, at each projected radius, the median velocity dispersion of models sampled in the Markov-Chain Monte Carlo analysis.  Dashed and dotted curves enclose the central 68\% and 95\% of velocity dispersion values from the sampled models.  The model profiles are fit to the unbinned kinematic data, but clearly show good agreement with the binned data plotted here. }
  \label{fig:dsphmassespar6_vprofiles}
  \end{center}
\end{figure*}

\begin{figure*}
\begin{center}
  \includegraphics[scale=0.9]{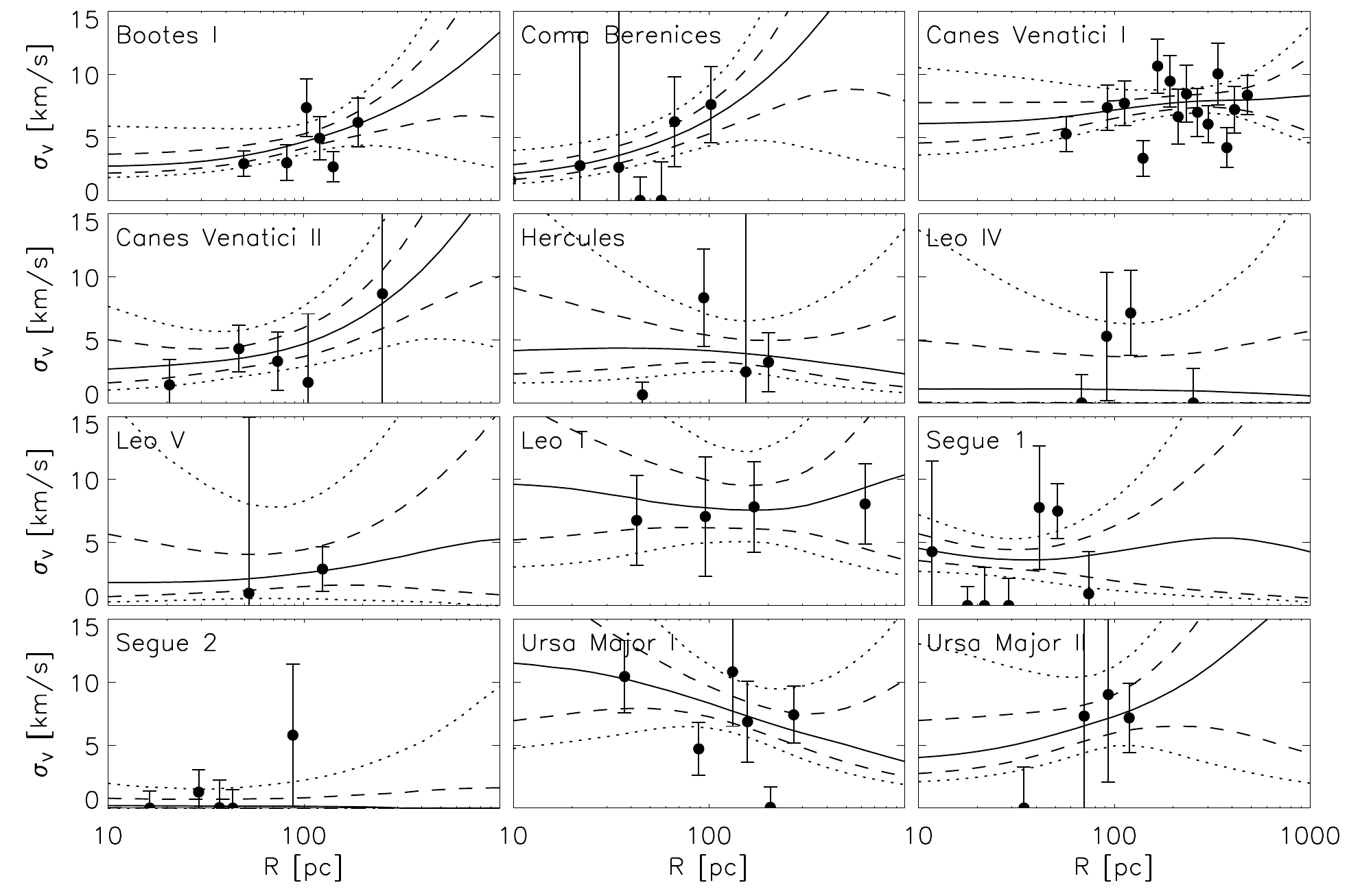}
  \caption{\scriptsize Same as Figure~\ref{fig:dsphmassespar6_vprofiles}, but for the Milky Way's ultra-faint satellites. In many bins the estimated velocity dispersion is zero because the actual dispersion is unresolved by the available data. As in Fig.~\ref{fig:dsphmassespar6_vprofiles} the points with error bars are for illustration; binned velocity dispersion estimates are not used in the fitting procedure.}
  \label{fig:2binufmassespar6_vprofiles}
  \end{center}
\end{figure*}

\subsection{Ultra-faint satellites}
\label{subsec:ultra-faint}

For the Milky Way's less luminous ``ultra-faint'' satellites discovered over the past seven years \citep[e.g.,][]{willman05a,zucker06a,belokurov07,belokurov08,belokurov09}, we make use of published data from a variety of sources.  We adopt projected halflight radii from the review of \citet{2012AJ....144....4M}.  Original sources are \citet{2008ApJ...684.1075M} and/or discovery papers for satellites discovered after that work \citep{belokurov08,belokurov09}.  

For Coma Berenices, Canes Venatici I, Canes Venatici II, Leo IV, Leo T, Ursa Major I and Ursa Major II, we utilize the Keck/Deimos velocity data of \citet{simon07}, generously provided by Marla Geha (private communication).  

For Hercules we adopt the stellar-kinematic data published by \citet{aden09b}.  After using Stromgren photometry to identify and remove foreground contamination independently of velocity, \citet{aden09} use these data to measure a global velocity dispersion of $3.7\pm 0.9$ km s$^{-1}$, slightly smaller than the earlier estimate of $5.1\pm 0.9$ km s$^{-1}$ by \citet{simon07}.  In the interest of deriving conservative estimates on the dark matter density (and ultimately the exclusion limits for particle models), we adopt the Hercules data of \citet{aden09b}.  

For Bo\"otes I, we adopt the stellar-kinematic data published by \citet{koposov11}.  The observing strategy of \citet{koposov11} provided $\sim 10$ independent velocity measurements of each star over the course of one month, thereby enabling a direct examination of intrinsic velocity variability (e.g., due to unresolved binary stars) that can potentially inflate observed velocity dispersions (and hence the inferred dark matter density) above the values attributable to the galaxy's gravitational potential \citep{mcconnachie10}.  \citet{koposov11} directly resolve velocity variability---including near-constant accelerations of $\sim 10$ km s$^{-1}$ month$^{-1}$---for $\sim 10\%$ of the stars in their sample.  In order to obtain conservative estimates of the dark matter density, we adopt the most restrictive sample of Bo\"otes I members identified by \citet[marked with a `B' in their Table 1]{koposov11}, which includes only the 37 stars that show no evidence for velocity variability and have small ($\leq 2.5$ km s$^{-1}$) velocity measurement errors.  

For Leo V, we adopt the stellar-kinematic data published by \citet{walker09c}.  These include seven stars identified as likely members.  However, while five of these stars lie within three times the projected half-light radius of Leo V ($r_h\sim 40$ pc; \citealt{belokurov08}), the other two lie $\ga 10r_h\sim 400$pc away from Leo V's center and in the direction toward Leo IV, which itself lies only $\sim 20$ kpc from Leo V.  This configuration is highly improbable for a dynamically relaxed system---even given the small number of stars---fueling speculation that Leo IV and Leo V are interacting gravitationally.  In that case, the two outermost stars in the Leo V sample may trace a stellar ``bridge'' of low surface brightness.  However, deep photometric studies have not yielded unambiguous evidence for such a structure \citep{dejong10,jin12,sand10,sand12}.  In any case, once again in the interest of conservatism, we consider only the five innermost members in the analysis of Leo V.  

For Segue 1, we adopt the stellar-kinematic data published by \citet{simon11}.  These data include repeat measurements for tens of stars with velocities originally measured by \citet{geha09}, enabling an analysis of velocity variability.  \citet{martinez11} perform a Bayesian analysis and conclude that the presence of unresolved binary stars is likely to have only mild (a $\sim 10\%$ effect) influence on estimates of Segue 1's intrinsic velocity dispersion.  Along with velocity measurements, we adopt the ``Bayesian'' membership probabilities listed for individual stars in Table 3 of \citet{simon11}. 

For Segue 2, we adopt the stellar-kinematic data published by \citet{2013ApJ...770...16K} for 25 member stars.  These measurements do not resolve Segue 2's internal velocity dispersion, instead placing a 95\% upper limit of $\sigma<2.6$ km s$^{-1}$.  A previous study by \citet{belokurov09} reported a velocity dispersion of $\sigma\sim 3.5$ km s$^{-1}$, based on a sample of $\sim 5$ member stars that they cautioned might be contaminated by members of a dynamically hotter stream in the same vicinity.  Again in the interest of placing conservative limits on the expected dark matter signal, we adopt the sample of \citet{2013ApJ...770...16K}---not only because it implies a smaller velocity dispersion, but also because it provides a larger number of member stars.  While the small velocity dispersion estimated by \citet{2013ApJ...770...16K} does not, by itself, require that Segue 2 is embedded within a dark matter halo, \citet{2013ApJ...770...16K} argue that the variance in metallicity among Segue 2's stars constitutes indirect evidence for a dark matter halo (whose deeper potential would help to retain chemically-enriched gas despite strong winds generated by star formation).

Finally, we note that while high-quality stellar-kinematic data sets are available for the Milky-Way satellites Sagittarius and Willman 1, these objects show strong evidence for tidal disruption and/or non-equilibrium kinematics \citep{1997AJ....113..634I,2011AJ....142..128W}.  Since any mass inference based on the Jeans equation relies fundamentally on the assumption of dynamic equilibrium, we do not consider these objects here.  

Table \ref{tab:dsphs} lists central coordinates, distances (from the Sun), absolute V-band magnitudes, and projected halflight radii for the Milky Way satellites we consider here.  The last column gives the number of member stars with velocity measurements available for the kinematic analysis.  Figures \ref{fig:dsphmassespar6_vprofiles} and \ref{fig:2binufmassespar6_vprofiles} display projected velocity dispersion profiles for each dwarf.  These binned profiles are included only for the purpose of display; the fitting that we describe below uses the unbinned data directly.

\begin{deluxetable*}{lrrrrrrrrrr}
  \tabletypesize{\scriptsize}
  \tablewidth{0pc}
  \tablecaption{Properties of Milky Way Satellites\tablenotemark{*} and Stellar-Kinematic Samples}
  \tablehead{\\
    \colhead{object}&\colhead{RA (J2000)}&\colhead{Dec. (J2000)}&\colhead{Distance}&\colhead{$M_V$}&\colhead{$R_{\rm half}$}&\colhead{$N_{\rm sample}$}&\colhead{$r_{\rm max}$}\\
    \colhead{}&\colhead{[hh:mm:ss]}&\colhead{[dd:mm:ss]}&\colhead{[kpc]}&\colhead{[mag]}&\colhead{[pc]}&\colhead{}&\colhead{[pc]}\\
  }
  \startdata
  \\
  Carina&06:41:36.7&$-$50:57:58&$105\pm 6$&$-9.1\pm 0.5$&$250\pm 39$&$774$    &  $2224^{+885}_{-441}$ \\[+0.1cm]
  Draco&17:20:12.4&+57:54:55&$76\pm 6$&$-8.8\pm 0.3$&$221\pm 19$&$292$        &  $1866^{+715}_{-317}$  \\[+0.1cm]

  Fornax&02:39:59.3&$-$34:26:57&$147\pm 12$&$-13.4\pm 0.3$&$710\pm 77$&$2483$ &  $6272^{+2616}_{-1366}$ \\[+0.1cm]

  Leo I&10:08:28.1&+12:18:23&$254\pm 15$&$-12.0\pm 0.3$&$251\pm 27$&$267$     &  $1948^{+794}_{-407}$ \\[+0.1cm]

  Leo II&11:13:28.8&+22:09:06&$233\pm 14$&$-9.8\pm 0.3$&$176\pm 42$&$126$     &  $824^{+345}_{-178}$  \\[+0.1cm]

  Sculptor&01:00:09.4&$-$33:42:33&$86\pm 6$&$-11.1\pm 0.5$&$283\pm 45$&$1365$ &  $2673^{+1099}_{-569}$  \\[+0.1cm]

  Sextans&10:13:03.0&$-$01:36:53&$86\pm 4$&$-9.3\pm 0.5$&$695\pm 44$&$441$    &  $2544^{+1109}_{-587}$  \\[+0.1cm]

  Ursa Minor&15:09:08.5&+67:13:21&$76\pm 3$&$-8.8\pm 0.5$&$181\pm 27$&$313$   &  $1580^{+626}_{-312}$ \\[+0.1cm]

  \\
  Bootes I&14:00:06.0&+14:30:00&$66\pm 2$&$-6.3\pm 0.2$&$242\pm 21$&$37$      &  $544^{+252}_{-135}$  \\[+0.1cm]

  Canes Venatici I&13:28:03.5&+33:33:21&$218\pm 10$&$-8.6\pm 0.2$&$564\pm 36$&$214$ &  $2030^{+884}_{-468}$   \\[+0.1cm]

  Canes Venatici II&12:57:10.0&+34:19:15&$160\pm 4$&$-4.9\pm 0.5$&$74\pm 14$&$25$   &  $352^{+105}_{-28}$  \\[+0.1cm]

  Coma Berenices&12:26:59.0&+23:54:15&$44\pm 4$&$-4.1\pm 0.5$&$77\pm 10$&$59$       &  $238^{+103}_{-53}$   \\[+0.1cm]

  Hercules&16:31:02.0&+12:47:30&$132\pm 12$&$-6.6\pm 0.4$&$330_{-52}^{+75}$&$30$    &  $638^{+295}_{-147}$  \\[+0.1cm]

  Leo IV&11:32:57.0&$-$00:32:00&$154\pm 6$&$-5.8\pm 0.4$&$206\pm 37$&$18$           &  $443^{+197}_{-95}$  \\[+0.1cm]

  Leo V&11:31:09.6&+02:13:12&$178\pm 10$&$-5.2\pm 0.4$&$135\pm 32$&$5$              &  $201^{+95}_{-43}$  \\[+0.1cm]

  Leo T&09:34:53.4&+17:03:05&$417\pm 19$&$-8.0\pm 0.5$&$120\pm 9$&$19$              &  $534^{+183}_{-60}$  \\[+0.1cm]

  Segue 1&10:07:04.0&+16:04:55&$23\pm 2$&$-1.5\pm 0.8$&$29_{-5}^{+8}$&$70$          &  $139^{+56}_{-28}$  \\[+0.1cm]

  Segue 2&02:19:16.0&+20:10:31&$35\pm 2$&$-2.5\pm 0.3$&$35\pm 3$&$25$                &  $119^{+45}_{-18}$  \\[+0.1cm]

  Ursa Major I&10:34:52.8&+51:55:12&$97\pm 4$&$-5.5\pm 0.3$&$319\pm 50$&$39$        &  $732^{+338}_{-181}$  \\[+0.1cm]

  Ursa Major II&08:51:30.0&+63:07:48&$32\pm 4$&$-4.2\pm 0.6$&$149\pm 21$&$20$       &  $294^{+139}_{-74}$  \\[+0.1cm]

  \enddata
  \tablenotetext{*}{Central coordinates, distances, absolute magnitudes and projected half-light radii are adopted from the review of \citet[see references to original sources therein]{2012AJ....144....4M}. }
  \label{tab:dsphs}
\end{deluxetable*}

\section{Fitting Procedure}
\label{sec:fitting}
Given the available kinematic data and adopted estimates of $R_e$ (which fixes $\Sigma(R)$ under the assumption of Plummer profiles), we fit models for $\rho(r)$ and $\beta_a(r)$ (see Sec.~\ref{sec:estimates}) following the procedure of \citet{2007PhRvD..75h3526S}.  Specifically, we assume that the velocity data sample a line-of-sight velocity distribution that is Gaussian\footnote{Given that we allow models with anisotropic and inherently non-Guassian velocity dispersions, this assumption of Gaussianity introduces an internal inconsistency.  However, by enabling the simple likelihood function given by Eq.~\eqref{eq:likelihood}, it avoids problems (e.g., arbitrariness of bin boundaries, unresolved dispersions) associated with analyses of binned profiles.  A more rigorous treatment would generate the likelihood function directly from a 6-D phase-space distribution function (M. Wilkinson, in preparation).}.  Thus we adopt the likelihood function
\begin{equation}
  L=\displaystyle\prod_{i=1}^{N}\, \,  \frac{1}{\left(2\pi\right)^{1/2} \left[ \delta^2_{u,i}+ \sigma^2(R_i) \right]^{1/2}}\exp\biggl [ -\frac{1}{2}\frac{(u_i-\langle u \rangle)^2}{\delta^2_{u,i}+ \sigma^2(R_i)}\biggr ],
  \label{eq:likelihood}
\end{equation}
where $u_i$ and $R_i$ are the line-of-sight velocity and magnitude of the projected position vector (with respect to the center of the dwarf) of the $i^{\rm th}$ star in the kinematic data set, $\delta_{u,i}$ is the observational error in the velocity, 
and $\sigma(R)$ is the velocity dispersion at projected position $R$, as specified by model parameters and calculated from Eq.~\eqref{eq:jeansproject}.  We consider only stars for which published probabilities of membership are greater than 0.95. The bulk velocity of the system $\langle u \rangle$ is a nuisance parameter that we marginalize over with a flat prior. Besides $\langle u \rangle$, the model has six free parameters and we adopt uniform priors (as in \citet{2011MNRAS.418.1526C}) over the following ranges:
\begin{itemize}
\item $-1\leq -\log_{10}[1-\beta_a]\leq +1$;
\item $-4\leq \log_{10}[\rho_s/(M_{\odot}\mathrm{pc}^{-3})]\leq +4$;
\item $0\leq \log_{10}[r_s/\mathrm{pc}]\leq +5$;
\item $0.5\leq \alpha\leq 3$;
\item $3\leq \beta\leq 10$;
\item $0\leq \gamma\leq 1.2$.
\end{itemize}
In order to sample the parameter space efficiently, we use the nested-sampling Monte Carlo algorithm introduced by \citet{2004AIPC..735..395S} and implemented in the software package MultiNest \citep{2008MNRAS.384..449F,2009MNRAS.398.1601F}, which outputs samples from the model's posterior probability distribution function (PDF).

\section{Physical considerations and truncation of halo profiles}
\label{sec:constraints}

\begin{figure*}
  \epsscale{1.1}
  \plotone{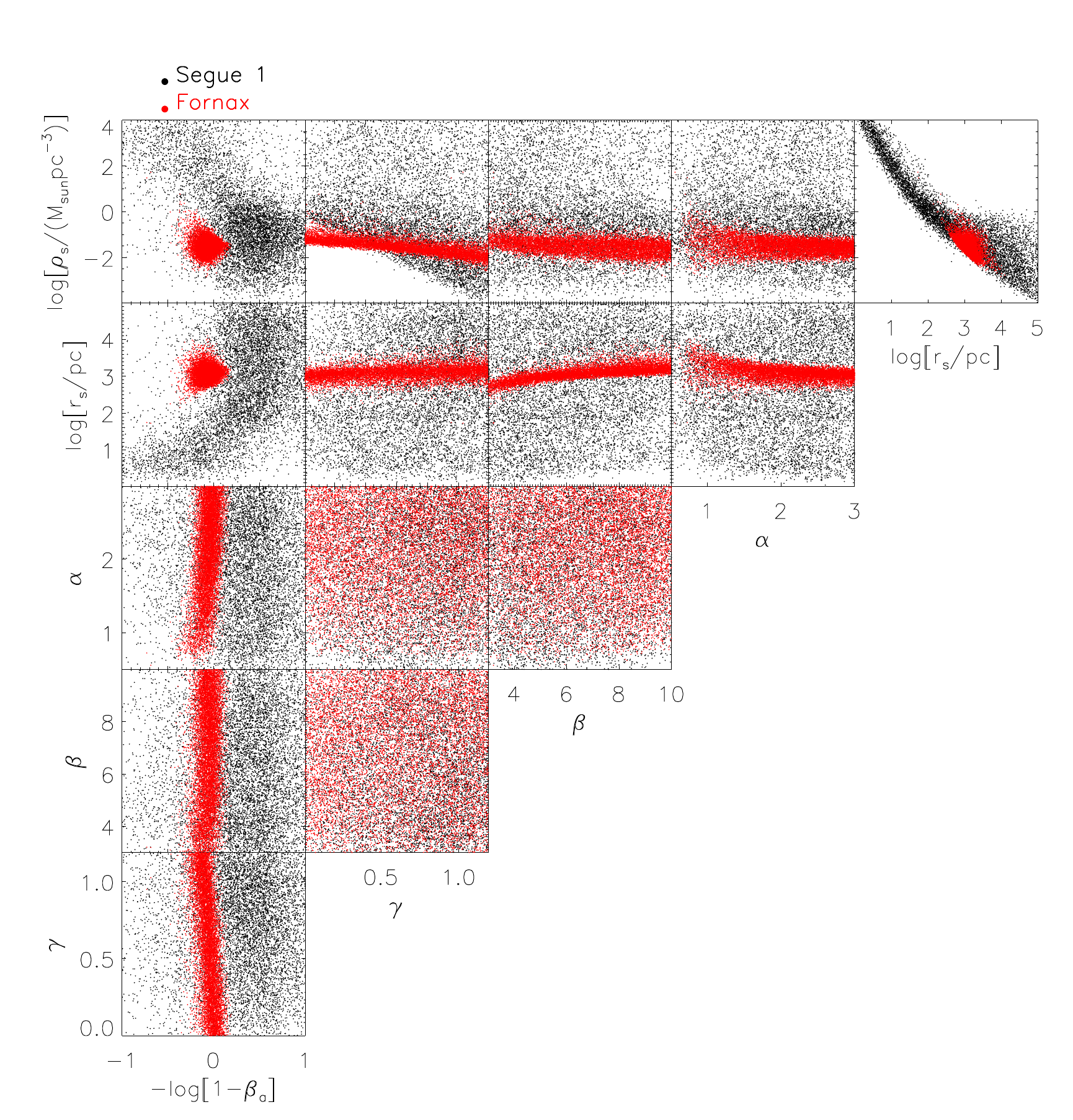}
  \caption{\scriptsize Samples from the posterior probability distribution functions of dark matter halo density profile and velocity anisotropy parameters for Segue 1 (black) and Fornax (red).}
\label{fig:mcmc}
\end{figure*}

Figure \ref{fig:mcmc} displays samples from the posterior PDFs returned by MultiNest for Fornax and Segue 1---the most luminous classical dwarf and one of the least luminous ultra-faints, respectively.   
As the model that we adopt for the halo density profile is free (and unconstrained by, e.g., N-body considerations) the kinematic data of each dwarf is compatible with a wide range of profiles. 
Therefore, we apply three additional filters to the kinematically-allowed dark matter density profiles.  The first two involve identifying an outer boundary for a given halo, while the third is a requirement that the halo formed in a cosmologically plausible way. 

\subsection{Halo truncation}

Given the form of Eq.~\eqref{eq:rho}, the annihilation rate will drop rapidly at galactocentric distances $r\gg r_s$ where the density profile is steeply falling ($\beta > 3$).  However, within $r_s$ the radial distribution of the emission is determined by the slope of the inner density profile $\gamma$.  For sufficiently cuspy profiles ($\gamma > 1$) the emission is dominated by annihilation near the halo center.   For $\gamma \sim 1$ the annihilation rate receives approximately equal contributions from all radii and for $\gamma < 1$ the emission comes primarily from the largest radii within $r_s$.

Unfortunately, the current data sets --- even for the classical dwarfs --- do not place strong upper bounds on $r_s$, thereby allowing emission that extends to an arbitrarily large radius\footnote{For spherically symmetric halos the mass exterior to a star's orbit exerts zero net force on that star \citep{1687pnpm.book.....N}; thus stellar kinematics in general carry no information about the mass distribution beyond the orbits of the stars.}. 
Therefore, the question of where the halo ends has important consequences for the expected dark matter signal from a dwarf galaxy.  The data for nearly all dwarf galaxies are consistent with density profiles described by single power laws with logarithmic slopes $d\log\rho/d\log r > -3$ --- indeed, despite its unphysically infinite mass, the ``isothermal sphere'', characterized by $\rho(r)\propto r^{-2}$, has long been used to model kinematics of spheroidal galaxies \citep{2008gady.book.....B}.  It is therefore important to define some means for preventing the outer parts of a halo --- i.e., regions outside the orbits of the observed stellar populations --- from dominating the integral used to calculate the $J$-profile (Eq.~\eqref{eqn:Jdef}).

\subsection{Truncating at the outermost observed star}
\label{subsec:outermost}

An obvious choice for a conservative truncation radius is that of the outermost member star used to estimate the velocity dispersion profile.  For stars well beyond the luminous scale radius $R_e$, the projected distances that we observe are likely to be similar to the de-projected distances $r$. However, if we observe enough stars close to the center it becomes likely that some of these stars lie at large galactocentric distances. Therefore, we use the entire distribution of projected radii of the kinematic sample to estimate the maximum galactocentric distance $r_{\mathrm{max}}$ among those stars.

We estimate $r_{\mathrm{max}}$ in the following way. Given spherical symmetry, it is straightforward to find the probability distribution for the unprojected distance to the outermost observed star given the projected distances to the observed stars.  We start by considering an individual star. Given its projected distance $R$ we take the probability of its line of sight distance $z$ (relative to the halo center) to be proportional to the (deprojected) Plummer density profile:
\begin{equation}
\Prob(z | R) \propto \left(1 + \frac{z^2 + R^2}{R_e^2}\right)^{-5/2},
\end{equation}
where $R_e$ is the projected halflight radius (Sec.~\ref{sec:estimates}). Once the above probability has been normalized (by integrating over $z$) we can construct the probability distribution for the unprojected distance $r$ given the projected distance $R$:
\begin{equation}
\Prob(r | R) = \int\limits_z \Prob(r|z,R) \,\, \Prob(z|R) \,dz.
\end{equation}
In the above $\Prob(r | z,R)$ is simply the Dirac delta function $\delta(r - \sqrt{z^2 + R^2})$. Note that the above integral is zero unless $r > R$, in which case the delta function picks out two values of z. The result we will need is the cumulative distribution function (CDF) of $r$ given $R$:
\begin{equation}
\CDF(r|R) = \int^r_0 \Prob(r' | R) dr' = \frac{\left(r^2 - R^2\right)^{1/2} \left(r^2 + \frac{1}{2}(3R_h^2 + R^2)\right)}{\left(r^2 + R_h^2\right)^{3/2}},
\label{eqn:CDFrgivenR}
\end{equation}
for $r>R$ and $\CDF(r|R) = 0$ for $r<R$.

To find the CDF for the distance to the outermost of $n$ observed stars we simply multiply the CDFs for each of the $n$ stars:
\begin{equation}
\CDF_{\rm max}(r | R_1, \dots, R_n) = \CDF(r|R_1) \cdots \CDF(r|R_n),
\label{eqn:CDFmax}
\end{equation}
where each term on the right-hand side is given by Eq.~\eqref{eqn:CDFrgivenR} and $R_i$ is the measured projected distance to the $i^{\rm th}$ member star.

For each dwarf, Eq.~\eqref{eqn:CDFmax} can be used to estimate the distance to the outermost star used in the Jeans analysis. The median estimate and $\pm 1 \sigma$ confidence intervals for the distance to the outermost member star for each dwarf are shown in the last column of Table \ref{tab:dsphs}. When computing $J$-profiles we truncate all halo profiles obtained from the Jeans/MultiNest sampling analysis at the median estimate to the outermost member star.

Note that this truncation is not imposed on the mass profile in Eq.~\eqref{eq:jeans2} when calculating the integral in Eq.~\eqref{eq:jeansproject} during the Jeans/MultiNest sampling.  However, the integral in Eq.~\eqref{eq:jeansproject} is dominated by the contribution from radii $r<R_e$, such that as long as the truncation radius is larger than the luminous effective radius (as it is for every dwarf galaxy we consider), the result from the Jeans/MultiNest analysis is insensitive to whether or not we truncate the halo density profile at the outermost star in the kinematic sample.  We have verified this argument by a controlled experiment and found that the significant effect of truncation is on the subsequent integration over the density profile that enters the calculation of the annihilation signal (see Eq.~\eqref{eqn:Jdef}).
For the purpose of this work (quantifying the expected dark matter flux from a dwarf) this particular choice of truncation is a conservative one.  A spherical Jeans analysis cannot, in principle, constrain the mass distribution far beyond the outermost member stars and we therefore set the density to zero at these distances.

\subsection{Tidal radius \label{sec:tidalradius}}

For some allowed models, however, the halo density at the galactocentric radius of the outermost star is far smaller than that expected for the Milky Way halo at the same location.  This situation would be inconsistent, as the outermost star (and dark matter particles) would likely be lost due to tides.  Therefore we impose an additional, physically-motivated filter by requiring that the tidal radius of any acceptable halo be larger than the distance to the outermost star.  

The magnitude of the tidal radius $r_t$ depends on the internal and external potentials (i.e., those of the dwarf and Milky Way, respectively), the orbit of the dwarf, and  on the orbital configuration within the dwarf (orbits that are prograde with respect to the dwarf's orbit are more easily stripped than those that are retrograde; \citealt{2006MNRAS.366..429R}).  

For each kinematically allowed halo profile we follow \citet{1957ApJ...125..451V} and \citet{1962AJ.....67..471K} and estimate a tidal radius $r_t$ by solving 
\begin{equation}
r_t^3 - D^3 \frac{M(r_t)}{ \MMW (D)} \left[ 2 + \frac{\omega^2 D^3}{G \MMW (D)} - \frac{d \ln \MMW}{d \ln r} \biggr\rvert_{r=D}  \right]^{-1} = 0.
\label{eq:tidalradius}
\end{equation}
Here, $D$ is the distance between the Milky Way center and the dwarf, $M(r_t)$ is the mass within the tidal radius of the dwarf galaxy, $\MMW(r)$ is the mass enclosed within radius $r$ in the Milky Way, and $\omega$ is the angular speed of the dwarf about the Galactic center (which we take to be a linear speed of 200 km/s divided by the distance from the dwarf to the Galactic center). The Milky Way mass model is taken to be an NFW profile with virial mass $\MMW = 10^{12} M_\odot$, a scale radius $r_{s,{\mathrm{MW}}} = 21.5 \, {\mathrm{kpc}}$, and a concentration $c_{\mathrm{MW}} = 12$~\citep{2002ApJ...573..597K} (for a more recent treatment of the subject see, e.g., \citealt{2013JCAP...07..016N}; the results are not sensitive to the uncertainties in the Milky Way mass model).

The expression in Eq.~\eqref{eq:tidalradius} should be taken as a crude approximation for several reasons (see, for example, discussions in \citet{2008gady.book.....B,2010gfe..book.....M}). First, it makes the assumption that the dwarf galaxy is on a circular orbit (tidal radii for systems with eccentric orbits are not well defined). Second, the three dimensional tidal surface is not of constant radius and thus does not correspond to one unique value for $r_t$. Third, Eq.~\eqref{eq:tidalradius} does not include the effect of orbital dynamics of the particles within the dwarf galaxy itself (manifested as some variance about the angular velocity $\omega$). Nevertheless, the utility of this prescription has been thoroughly explored in studies of globular clusters, dark matter substructure, and dwarf galaxies \citep{1998ApJ...495..297J, 2001ApJ...557..137J,2001ApJ...559..716T, 2004MNRAS.348..811T,2002PhRvD..66d3003Z,2003ApJ...598...49Z}, and gives a reasonably intuitive definition of ``tidal radius''.  

Therefore, we reject any halo profiles for which this estimate of the tidal radius is smaller than the radius we estimate for the outermost member in the kinematic samples\footnote{We use the minimum possible distance to the outermost star: the largest projected distance to any of the members. This has a conservative impact in the resulting $J$ values.}. This consistency condition turns out to affect only two dwarfs, the ultra-faints Segue 2 and Leo IV, removing profiles with small values of both $\rho_s$ and $r_s$ (i.e. profiles that do not lie on the typical $\rho_s$ vs. $r_s$ constraint seen in Fig.~\ref{fig:mcmc}).

\subsection{Cosmological considerations \label{sec:cosmo}}

As a final filter, we reject halo profiles that would be cosmologically ``implausible'' in the sense that their formation would have required extremely rare peaks in the primordial matter density field.   
We estimate the rareness of a candidate halo in the following way. Cosmological simulations show that the density profile near the center of a dark matter halo is more or less set at the time of its formation (modulo feedback effects from baryon-physical processes).  The outer profile then evolves as mass is accumulated by accretion in a hierarchical scenario (see e.g., \citet{1997ApJ...490..493N,2006ApJ...652...71W,2008MNRAS.387..536G}). To a first approximation, then, the primordial mass of a halo (the virial mass of the collapsing overdensity) is greater (or on the order of) the mass $M$ within $r_s$ today.
In addition, the density of the inner parts of a halo is also roughly set at the time of formation, following a distribution in concentration, $c\equiv R_{\rm vir}/r_s$, where $R_{\rm vir}$ is the radius of the virialized region \citep{1996ApJ...462..563N}.

For example, the scale density of an NFW profile is $\rho_s = \delta_c \rho_{\rm crit} f(c_{\mathrm{v}})$, where $\delta_c$ is the characteristic overdensity of a collapsed object and $\rho_{\rm crit}$ (scaling as $(1+z)^3$ in the matter-dominated era) is the critical density of the Universe at the time of collapse~\citep[e.g.][]{2008MNRAS.387..536G}. The function $f(c_{\mathrm v}) = \ln( 1+ \cv ) + \cv / ( 1 + \cv) $ is a function  of order unity. There is thus an approximately one-to-one mapping between the collapse redshift and the scale density $\rho_s$ of a halo (ignoring the intrinsic variation in concentration \cite{2006ApJ...652...71W,2008MNRAS.387..536G}). For simplicity, and due to a lack of knowledge of the initial conditions that give rise to the intrinsic properties of the dwarf galaxies at their formation time, we use the approximation  \citep{1996ApJ...462..563N}
\begin{equation}
\frac{1 + z_{\rm col}}{20} \gtrsim  \left(\frac{\rho_s}{M_\odot{\rm pc}^{-3}}\right)^{1/3}.
\label{eqn:zcol}
\end{equation}
as an {\it upper}  limit on the value of the characteristic density of each halo in the chains.

The collapse redshift, along with an estimate of a halo's mass at collapse $M$, can be used to quantify the rarity of the density perturbation that collapsed to form the halo. In order to estimate the mass of the collapsed perturbation we use the conventional theory of spherical collapse. A region of space has collapsed and virialized when its average overdensity (with respect to the background as computed in linear perturbation theory) is $\delta_{\rm collapse}\sim 1.686$. The overdensity at redshift $z$, averaged over a region of mass $M$, is a Gaussian random variable with mean 0 and standard deviation $\sigma(M) {\cal{D}}(z)$, where $\sigma(M) = \langle \delta_M^2 \rangle^{1/2}$ is the standard deviation of the overdensity today when smoothed over a region of mass $M$ and ${\cal{D}}(z)$, the growth function, quantifies the linear growth of perturbations such that ${\cal{D}}(z=0)=1$. Therefore, the rarity (in standard deviations) of a region of mass $M$ collapsing at redshift $z$ is given by (see e.g.~\citet{1991ApJ...379..440B,1991MNRAS.248..332B,1993MNRAS.262..627L}),
\begin{equation}
\nu(M,z) = \frac{\delta_{\rm collapse}}{\sigma(M) {\cal{D}}(z)}.
\label{eqn:nudef}
\end{equation}

The probability that such a region has collapsed by redshift $z$ is the tail probability of a standard normal distribution to the right of $\nu$. We can make a rough estimate of the probability that the Milky Way contains {\em any} objects of mass $M$ which formed at (or before) a redshift $z$ by incorporating a trials factor which counts the number of ``independent'' regions of the Universe of mass $M$ that eventually make up the Milky Way today. We estimate this quantity simply as $\MMW / M$, where $\MMW = 10^{12} M_\odot$ is the mass of the Milky Way. We take the mass $M$ of a halo to be the lesser of the mass within $r_s$ and the mass within the outermost star. For the halos generated by the Monte Carlo sampling the masses range from about $10^{5}$ to $10^{7} M_\odot$. This leads to trials factors between $10^5$ and $10^7$. The probability that the Milky Way contains a halo that collapsed with mass $M$ by redshift $z$ is then
\begin{equation}
\Prob(M,z) = 1 - \Phi\left[\nu(M,z)\right]^{\MMW/M}, 
\label{eqn:PMz}
\end{equation}
where $\Phi(x)$ is the cumulative distribution function of a standard normal distribution and $\nu(M,z)$ is given by Eq.~\eqref{eqn:nudef}. We apply the constraint by demanding that the Milky Way is ``typical'' at the $3\sigma$ level, i.e. we reject a halo if $\Prob(M,z)  < 0.003$.
For a trials factor of $10^6$ this constraint is equivalent to demanding that $\nu \lesssim 6$. The results are insensitive to the threshold value for $\Prob(M,z)$ --- the probability of the Milky Way containing a halo goes to zero extremely rapidly as the halo's $\nu$ value increases beyond about $\nu \approx 6$.

In applying this cosmological argument {\it a posteriori} on the chains we find that in the case of classical dwarfs (for which we have a large number of gravitational stellar tracers) stellar kinematics alone does not allow for rare peaks and this filter essentially has no effect. However for the ultra-faint dwarfs the stellar kinematic data allow dark matter halos that appear to be extremely rare, suggesting that the origin of such rare peaks is due to the small size of kinematic samples available compared with the classical dwarfs. In practice, the halos which are eliminated are those with $\rho_s \gtrsim 1 \, M_\odot /\mathrm{pc}^{3}$.

\section{Results}
\label{sec:results}

\begin{figure*}
  \epsscale{1.2}
  \plotone{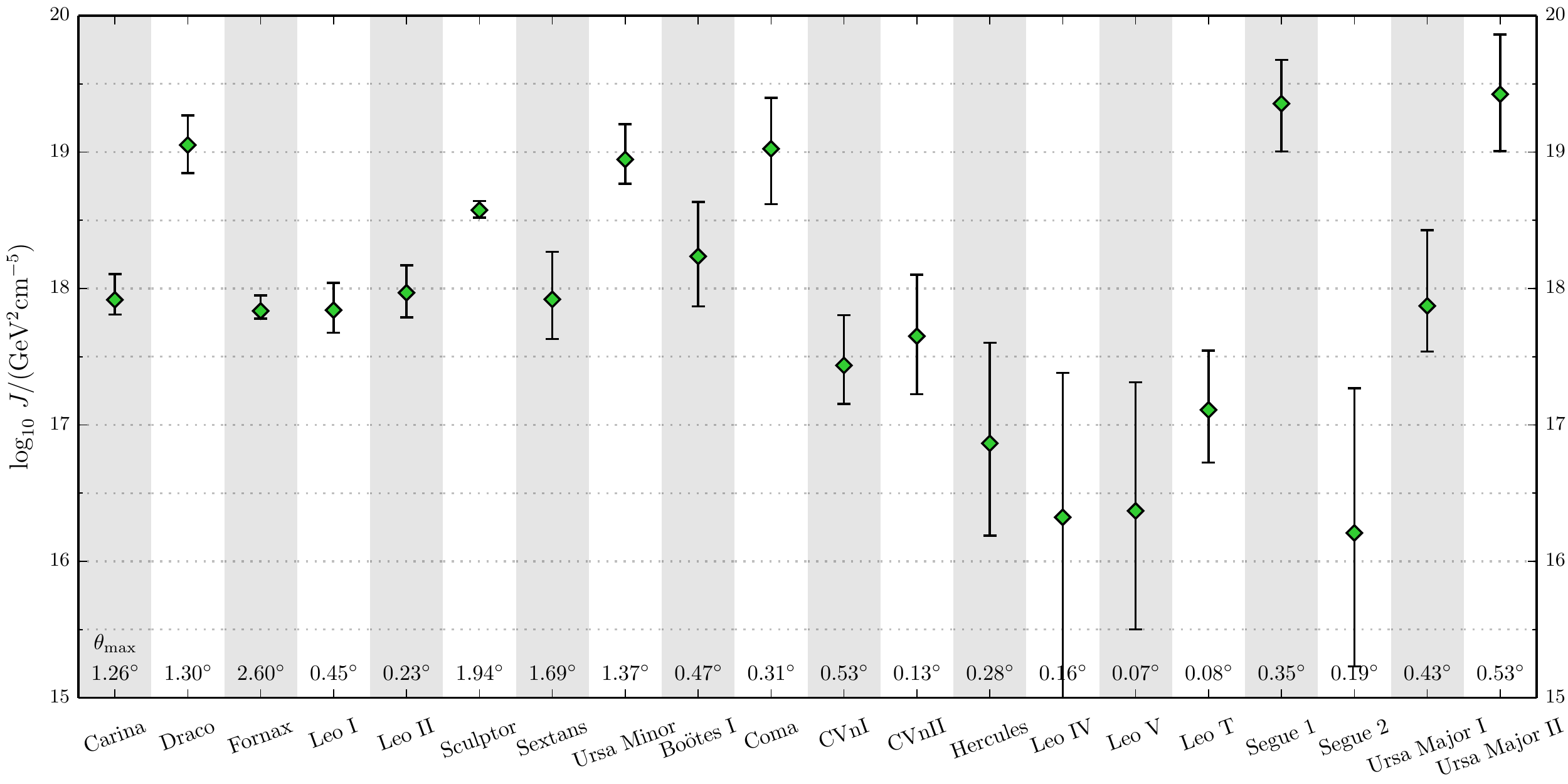}
  \caption{Annihilation $\J$-profiles integrated out to $\thetamax$ for all dwarf galaxies. The error bars show the $1\sigma$ allowed range in the value of $\J$ based on the kinematic analysis described in the text. The most prominent dwarf galaxies for annihilation studies are Draco and Ursa Minor (classical) and Coma Berenices, Segue 1, and Ursa Major II (ultra-faint). Numerical values are provided in Table~\ref{tab:Jtable} and in machine-readable form as described in Appendix~\ref{sec:appendix}.}
  \label{fig:fig_Jtot_all}
\end{figure*}
\begin{figure*}
  \epsscale{1.2}
  \plotone{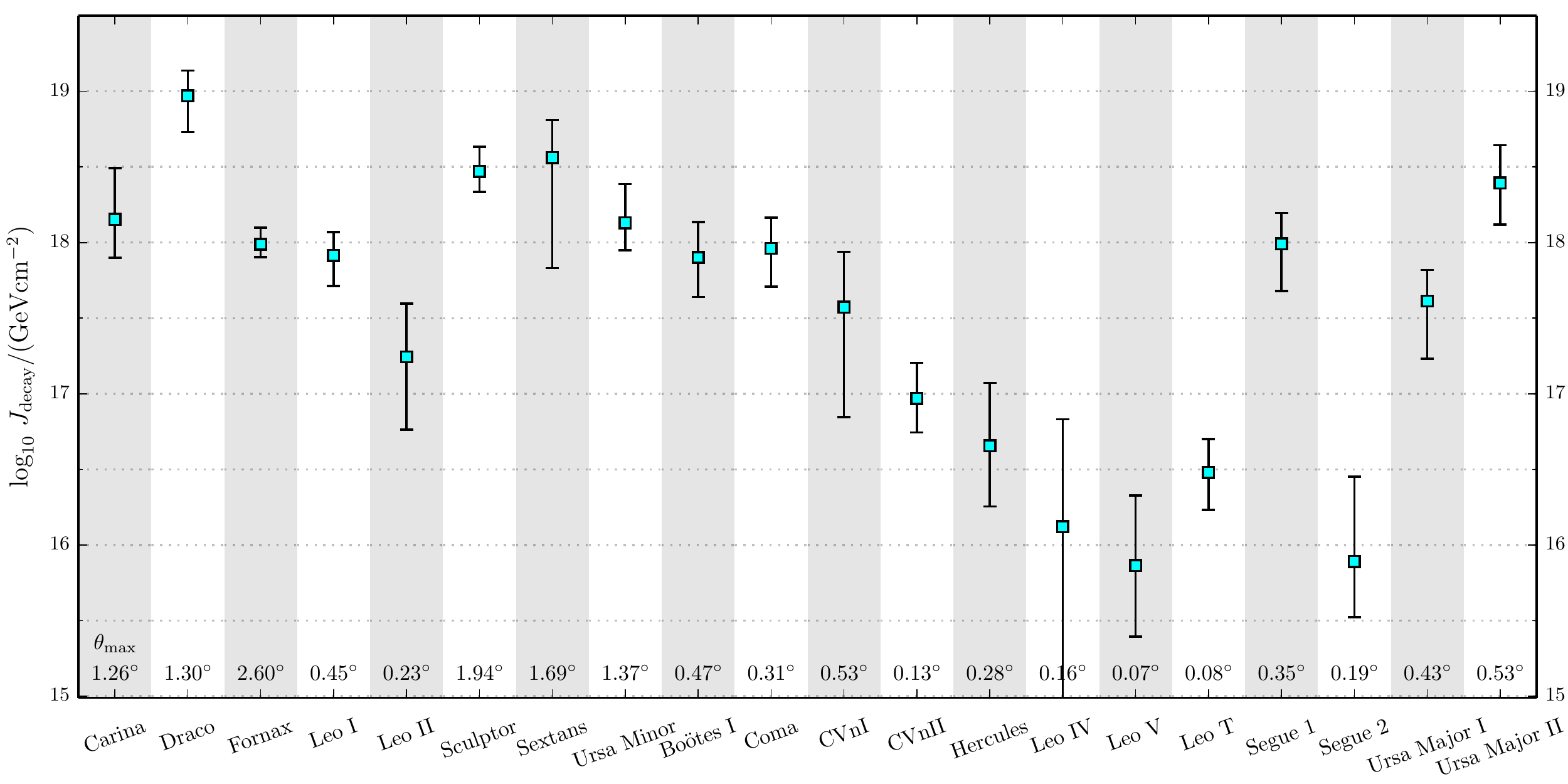}
  \caption{Same as Fig.~\ref{fig:fig_Jtot_all} for but for dark matter decay ($J_{\rm decay}$). The most prominent dwarf galaxy for decay studies is Draco, with Sculptor and Sextans both a factor of about 3 fainter. Numerical values are provided in Table~\ref{tab:Jtable} and in machine-readable form as described in Appendix~\ref{sec:appendix}.}
  \label{fig:fig_Jdecaytot_all}
\end{figure*}

\subsection{$J$-profiles for annihilation and decay} 

The gamma-ray flux from dark matter annihilation is fully described by the function $dJ(\theta)/d\Omega$ and any dark matter search is properly conducted using this function (e.g.~\citet{2014arXiv1410.2242G}). However, it is useful to explore the dark matter distributions in dwarfs using summary quantities based on the $J$-profiles.

Concerning indirect detection, the most important properties of the dwarf galaxies are the amplitude and spatial extent of the $J$-profile. The amplitude is often given as the integral of $dJ(\theta)/d\Omega$ over some solid angle; i.e. we define
\begin{equation}
J(\theta) \equiv \int_0^\theta \frac{dJ(\theta')}{d\Omega} 2\pi \sin(\theta')d\theta',
\label{eqn:defJtheta}
\end{equation}
with an analogous definition for $J_{\rm decay}(\theta)$. A detector-independent amplitude is the $J$-profile integrated out to the truncation radius of the halo. This corresponds to integrating $dJ(\theta)/d\Omega$ out to an angle $\theta_\mathrm{max} = \mathrm{arcsin}(r_\mathrm{max}/D)$, where $r_\mathrm{max}$ is the distance from the center of the dwarf to the outermost member star (Sec.~\ref{subsec:outermost}) and $D$ is the distance from Earth to the dwarf.  Integrating the $J$-profile within $\theta_\mathrm{max}$ gives the {\it total dark matter flux expected from a halo}. We use $J$ or $J_{\rm decay}$ to denote this scalar quantity when there is no confusion.

\begin{deluxetable*}{lccccccccc}
  \tabletypesize{\scriptsize}
  \tablewidth{0pc}
  \tablecaption{$J$ values for annihilation and decay in dwarf galaxies\tablenotemark{*}}
  \tablehead{\\
  {Dwarf}	&	{$\thetamax$}	&	{$\theta_{0.5}$}	&	{$\theta_{0.5\, \rm decay}$}	&	{$\log_{10} J(\thetamax)$}	&	{$\log_{10}J(0.5 \degg)$}	&	{$\log_{10}\Jdecay(\thetamax)$}	&	{$\log_{10} \Jdecay(0.5 \degg)$} \\
    \colhead{}	&	\colhead{$[\deg]$}	&	\colhead{$[\deg]$}	&	\colhead{$[\deg]$}	& \colhead{${\mathrm{[GeV^2 cm^{-5}]}}$}	&	\colhead{${\mathrm{[GeV^2 cm^{-5}]}}$}	&	\colhead{${\mathrm{[GeV cm^{-2}]}}$}	&	\colhead{${\mathrm{[GeV cm^{-2}]}}$}}\\
  \startdata
  \\
Carina         &  $1.26$  &  $0.15^{+0.15}_{-0.07}$  &  $0.46^{+0.16}_{-0.12}$  &  $17.92^{+0.19}_{-0.11}$  &  $17.87^{+0.10}_{-0.09}$  &  $18.15^{+0.34}_{-0.25}$  &  $17.90^{+0.17}_{-0.16}$  &     \\[+0.1cm]
Draco          &  $1.30$  &  $0.40^{+0.16}_{-0.15}$  &  $0.64^{+0.06}_{-0.14}$  &  $19.05^{+0.22}_{-0.21}$  &  $18.84^{+0.12}_{-0.13}$  &  $18.97^{+0.17}_{-0.24}$  &  $18.53^{+0.10}_{-0.12}$  &     \\[+0.1cm]
Fornax         &  $2.61$  &  $0.13^{+0.04}_{-0.05}$  &  $0.31^{+0.08}_{-0.05}$  &  $17.84^{+0.11}_{-0.06}$  &  $17.83^{+0.12}_{-0.06}$  &  $17.99^{+0.11}_{-0.08}$  &  $17.86^{+0.04}_{-0.05}$  &     \\[+0.1cm]
Leo I          &  $0.45$  &  $0.13^{+0.05}_{-0.05}$  &  $0.22^{+0.02}_{-0.04}$  &  $17.84^{+0.20}_{-0.16}$  &  $17.84^{+0.20}_{-0.16}$  &  $17.91^{+0.15}_{-0.20}$  &  $17.91^{+0.15}_{-0.20}$  &     \\[+0.1cm]
Leo II         &  $0.23$  &  $0.04^{+0.05}_{-0.02}$  &  $0.09^{+0.03}_{-0.05}$  &  $17.97^{+0.20}_{-0.18}$  &  $17.97^{+0.20}_{-0.18}$  &  $17.24^{+0.35}_{-0.48}$  &  $17.24^{+0.35}_{-0.48}$  &     \\[+0.1cm]
Sculptor       &  $1.94$  &  $0.15^{+0.05}_{-0.05}$  &  $0.48^{+0.14}_{-0.11}$  &  $18.57^{+0.07}_{-0.05}$  &  $18.54^{+0.06}_{-0.05}$  &  $18.47^{+0.16}_{-0.14}$  &  $18.19^{+0.07}_{-0.06}$  &     \\[+0.1cm]
Sextans        &  $1.70$  &  $0.58^{+0.32}_{-0.47}$  &  $0.87^{+0.10}_{-0.53}$  &  $17.92^{+0.35}_{-0.29}$  &  $17.52^{+0.28}_{-0.18}$  &  $18.56^{+0.25}_{-0.73}$  &  $17.89^{+0.13}_{-0.23}$  &     \\[+0.1cm]
Ursa Minor     &  $1.37$  &  $0.06^{+0.07}_{-0.03}$  &  $0.25^{+0.14}_{-0.09}$  &  $18.95^{+0.26}_{-0.18}$  &  $18.93^{+0.27}_{-0.19}$  &  $18.13^{+0.26}_{-0.18}$  &  $18.03^{+0.16}_{-0.13}$  &     \\[+0.1cm]
\\
Bo{\"o}tes I   &  $0.47$  &  $0.22^{+0.05}_{-0.10}$  &  $0.26^{+0.02}_{-0.04}$  &  $18.24^{+0.40}_{-0.37}$  &  $18.24^{+0.40}_{-0.37}$  &  $17.90^{+0.23}_{-0.26}$  &  $17.90^{+0.23}_{-0.26}$  &     \\[+0.1cm]
Coma           &  $0.31$  &  $0.16^{+0.02}_{-0.05}$  &  $0.17^{+0.01}_{-0.02}$  &  $19.02^{+0.37}_{-0.41}$  &  $19.02^{+0.37}_{-0.41}$  &  $17.96^{+0.20}_{-0.25}$  &  $17.96^{+0.20}_{-0.25}$  &     \\[+0.1cm]
CVnI           &  $0.53$  &  $0.11^{+0.15}_{-0.09}$  &  $0.23^{+0.07}_{-0.17}$  &  $17.44^{+0.37}_{-0.28}$  &  $17.43^{+0.37}_{-0.28}$  &  $17.57^{+0.37}_{-0.73}$  &  $17.57^{+0.36}_{-0.72}$  &     \\[+0.1cm]
CVnII          &  $0.13$  &  $0.07^{+0.01}_{-0.02}$  &  $0.07^{+0.00}_{-0.01}$  &  $17.65^{+0.45}_{-0.43}$  &  $17.65^{+0.45}_{-0.43}$  &  $16.97^{+0.24}_{-0.23}$  &  $16.97^{+0.24}_{-0.23}$  &     \\[+0.1cm]
Hercules       &  $0.28$  &  $0.07^{+0.08}_{-0.06}$  &  $0.12^{+0.03}_{-0.09}$  &  $16.86^{+0.74}_{-0.68}$  &  $16.86^{+0.74}_{-0.68}$  &  $16.66^{+0.42}_{-0.40}$  &  $16.66^{+0.42}_{-0.40}$  &     \\[+0.1cm]
Leo IV         &  $0.16$  &  $0.05^{+0.03}_{-0.04}$  &  $0.08^{+0.01}_{-0.06}$  &  $16.32^{+1.06}_{-1.69}$  &  $16.32^{+1.06}_{-1.69}$  &  $16.12^{+0.71}_{-1.14}$  &  $16.12^{+0.71}_{-1.14}$  &     \\[+0.1cm]
Leo V          &  $0.07$  &  $0.03^{+0.01}_{-0.02}$  &  $0.04^{+0.00}_{-0.01}$  &  $16.37^{+0.94}_{-0.87}$  &  $16.37^{+0.94}_{-0.87}$  &  $15.86^{+0.46}_{-0.47}$  &  $15.86^{+0.46}_{-0.47}$  &     \\[+0.1cm]
Leo T          &  $0.08$  &  $0.03^{+0.01}_{-0.02}$  &  $0.04^{+0.00}_{-0.01}$  &  $17.11^{+0.44}_{-0.39}$  &  $17.11^{+0.44}_{-0.39}$  &  $16.48^{+0.22}_{-0.25}$  &  $16.48^{+0.22}_{-0.25}$  &     \\[+0.1cm]
Segue 1        &  $0.35$  &  $0.13^{+0.05}_{-0.07}$  &  $0.18^{+0.01}_{-0.05}$  &  $19.36^{+0.32}_{-0.35}$  &  $19.36^{+0.32}_{-0.35}$  &  $17.99^{+0.20}_{-0.31}$  &  $17.99^{+0.20}_{-0.31}$  &     \\[+0.1cm]
Segue 2        &  $0.19$  &  $0.07^{+0.03}_{-0.05}$  &  $0.10^{+0.01}_{-0.05}$  &  $16.21^{+1.06}_{-0.98}$  &  $16.21^{+1.06}_{-0.98}$  &  $15.89^{+0.56}_{-0.37}$  &  $15.89^{+0.56}_{-0.37}$  &     \\[+0.1cm]
Ursa Major I   &  $0.43$  &  $0.15^{+0.08}_{-0.12}$  &  $0.22^{+0.03}_{-0.14}$  &  $17.87^{+0.56}_{-0.33}$  &  $17.87^{+0.56}_{-0.33}$  &  $17.61^{+0.20}_{-0.38}$  &  $17.61^{+0.20}_{-0.38}$  &     \\[+0.1cm]
Ursa Major II  &  $0.53$  &  $0.24^{+0.06}_{-0.11}$  &  $0.29^{+0.02}_{-0.04}$  &  $19.42^{+0.44}_{-0.42}$  &  $19.42^{+0.44}_{-0.42}$  &  $18.39^{+0.25}_{-0.27}$  &  $18.38^{+0.25}_{-0.27}$  &     \\[+0.1cm]
\enddata
  \tablenotetext{*}{Errors correspond to the 1$\sigma$ range of the kinematic analysis (i.e. the 16$^{\rm th}$ and 84$^{\rm th}$ percentiles). $J(\theta)$ is the $J$-profile (Eq.~\eqref{eqn:Jdef}) integrated over a cone with radius $\theta$ (Eq.~\eqref{eqn:defJtheta}); $\theta_{0.5}$ is the ``half-light radius'' for dark matter emission (i.e. the angle containing 50\% of the total emission: $J(\theta_{0.5}) = 0.5 \times J(\thetamax)$, see Sec.~\ref{subsec:spatialextent}). Analogous definitions apply for dark matter decay. Halos are truncated at a radius corresponding to $\thetamax$ (Sec.~\ref{subsec:outermost}).}
  \label{tab:Jtable}
\end{deluxetable*}

Using the kinematic data discussed in Sec.~\ref{sec:observations}, and applying the formalism described in Secs.~\ref{sec:estimates}, \ref{sec:fitting}, and \ref{sec:constraints}, we apply equations~\eqref{eqn:Jdef} and~\eqref{eqn:Jdecaydef} to estimate the kinematically allowed values of $J$ and $\Jdecay$. Figures~\ref{fig:fig_Jtot_all} and \ref{fig:fig_Jdecaytot_all} along with Table~\ref{tab:Jtable} show the main results of this work.

Figure~\ref{fig:fig_Jtot_all} shows the derived $\J$ values for all of the dwarf galaxies considered in this analysis. Error bars represent the 1$\sigma$ uncertainty in the value of $J$ (e.g. the 16$^{\rm th}$ and 84$^{\rm th}$ percentiles of the posterior distriution). There are many interesting features of this distribution of $\J$ values among the dwarf galaxies. Concerning the overall amplitude of an annihilation signal, we find that among the classical dwarfs Draco and Ursa Minor have the largest expected flux. However, Sculptor and, to a lesser extent, Fornax and Carina have the most constraining kinematic sample, giving a very small range of allowed values for $J$ and thus the smallest uncertainties (in agreement with the analysis of \citealt{2011MNRAS.418.1526C}). Among the ultra-faint dwarf galaxies, Segue 1 and Ursa Major II have central $J$ values that are higher then those of any of the classical dwarfs. However, the uncertainties in these values are also larger than those of all the classical dwarfs. Of all the dwarf galaxies in this sample, Sculptor has the smallest uncertainty in its total $J$ value (about 15\%). On the other hand, all ultra-faint dwarf galaxies exhibit 1$\sigma$ errors that are about an order of magnitude (in some cases several orders of magnitude -- e.g., Leo IV). This is an outcome of the limited size of the kinematic samples available to date. From the entire ultra-faint sample, Segue 1, Coma Berenices, and Ursa Major II exhibit well-constrained (less than an order of magnitude) $J$ values, and due to their overall high amplitude they should be considered for current and future annihilation searches.

\begin{figure}
\includegraphics[scale=1]{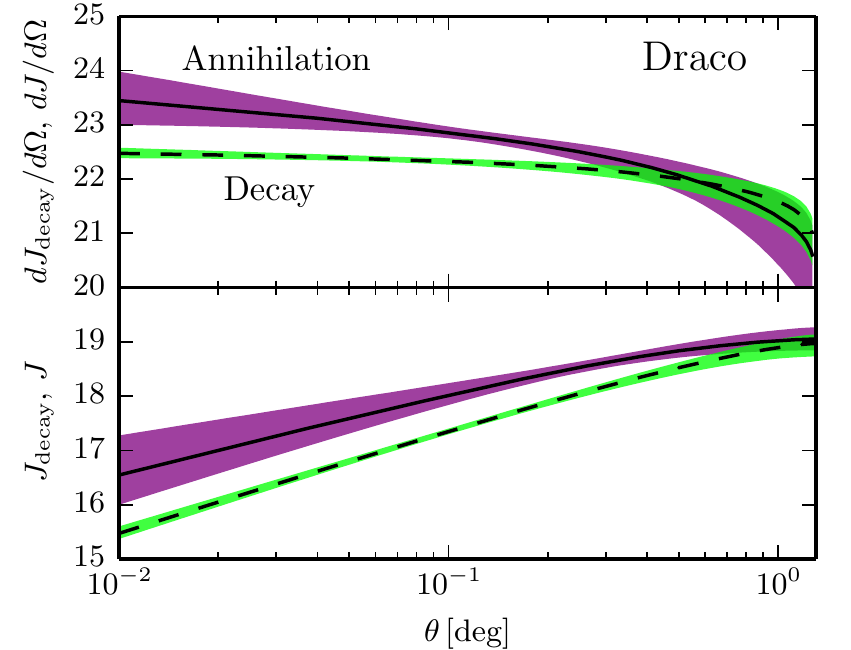}\\
\includegraphics[scale=1]{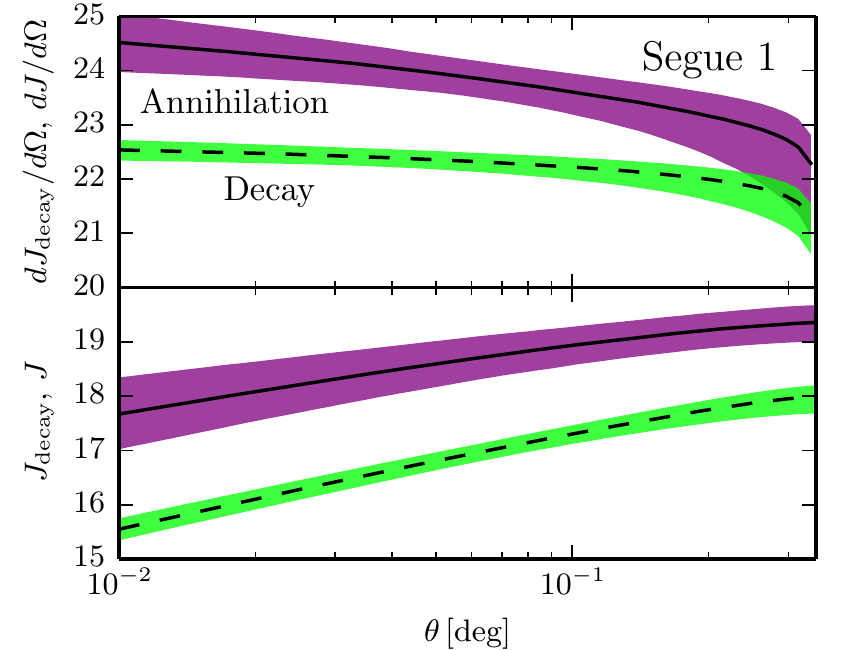}
  \caption{Expected emission profiles for annihilation (purple) and decay (green) for Draco and Segue 1. At each angle the solid and dashed lines show the median profiles and the shaded band corresponds to the $\pm 1\sigma$ distribution as derived in the kinematic analysis. The top panels show $\log_{10}dJ(\theta)/d\Omega$ (purple), and $\log_{10}\Jdecay(\theta)/d\Omega$ (green) (see Eqs.~\eqref{eqn:Jdef} and~\eqref{eqn:Jdecaydef}) in units of $\mathrm{GeV^2 \,cm^{-5}}$ and $\mathrm{GeV\, cm^{-2}}$ respectively. The lower panels show these quantities integrated over a solid angle of radius $\theta$ (Eq.~\eqref{eqn:defJtheta}). These envelopes should be thought of as giving the uncertainty in the $J$-profile and integrated $J$-profile at each value of $\theta$. (Integrated $J$ vs. $\theta$ and $J_{\rm decay}$ vs. $\theta$ constraints for all the dwarfs are available in machine-readable form as described in Appendix~\ref{sec:appendix}.)}
  \label{fig:fig_profiles_Draco_Seg1}
\end{figure}

Figure~\ref{fig:fig_Jdecaytot_all} shows the corresponding $\Jdecay$ values for the twenty dwarf galaxies analyzed. As in Fig.~\ref{fig:fig_Jtot_all}, error bars represent the 1$\sigma$ range in the value of $\Jdecay$. In the case of dark matter decay scenarios, the emission profile is set by the amount of mass along the line of sight (Eq.~\eqref{eqn:dF_decay}) and is less sensitive to the inner slope of the density profile. As a result, the preferential ordering of dwarfs according to their emission amplitude is different than when searching for annihilation products. In the case of decay, the kinematic analysis shows that Draco, and to a lesser extent, Sextans and Sculptor are the sources with the largest expected emission. As in the annihilation case, Fornax and Sculptor have the least systematic uncertainty in their emission amplitudes compared with the rest of the dwarfs.

Many of the ultra-faint dwarfs do not appear to be as promising targets as several of the classical dwarfs. This is due to the conservative way we chose to truncate the halos when computing $\J$ and $\Jdecay$. The outermost observed member stars of the ultra-faints are much closer in than those in the classical dwarfs  (see the $\theta_{\rm max}$ values along the bottom of Figs.~\ref{fig:fig_Jtot_all} and~\ref{fig:fig_Jdecaytot_all} and the last column of Table~\ref{tab:dsphs}). Therefore, we simply do not allow the ultra-faint halos to be as extended as those of the classical dwarfs. Since the emission profile from decay is more sensitive to the total mass of the halo, whereas the annihilation profile is more sensitive to the inner slope, the effect of the truncation has different effects when considering annihilation and decay. Whether this truncation is conservative or if it truly reflects the physical sizes of the ultra-faints' dark matter halos is unknown.

It is important to note that the ranking of the various $J$ values implies a consistency check for any detection claim. First, it is likely that if a signal is seen, it will be seen in multiple dwarfs. Consider the pairs Draco/Ursa Minor (classical) and Segue 1/Ursa Major II (ultra-faint). The similar $J$ values among the members of a pair imply that if an annihilation signal is detected in any of these dwarfs there should be a detection of similar amplitude in the other member of the pair (modulo differences in the diffuse gamma-ray background between the dwarfs).

In the farther future, this argument may be turned around to provide an example of ``dark matter particle astronomy'': the relative annihilation fluxes measured in multiple dwarfs can be used to constrain the dark matter distributions in their halos. For example, the detection of a signal in a highly constrained dwarf like Sculptor, Draco, or Ursa Minor would immediately tell us something about the dark matter distribution in Segue 1, an object less luminous by 3 orders of magnitude.

\subsection{Spatial extent}
\label{subsec:spatialextent}

The question of whether a dark matter halo will appear as an extended source for a gamma-ray instrument is important as the additional information available from the angular distribution of emission can be used to increase the signal-to-noise ratio. More point-like sources of emission are more straightforward to detect. However, the detection of a spatially varying annihilation signal immediately reveals properties of the dark matter halo.

Figure~\ref{fig:fig_profiles_Draco_Seg1} shows the constraints on the emission profiles for two benchmark dwarf galaxies: the classical dwarf Draco and the ultra-faint dwarf Segue 1. The envelopes show the $\pm 1\sigma$ and median values of the emission profiles as a function of the angular separation from the center of the dwarf. It is important to emphasize that none of the curves in the figure correspond to an individual dark matter profile. Rather, the envelope should be thought of as constraining the value of $dJ/d\Omega$ or $d\Jdecay/d\Omega$ at a given angular separation. In the lower panels we show the emission profiles integrated over solid angle out to an angular separation $\theta$ (Eq.~\eqref{eqn:defJtheta}). For Draco we see a familiar result: there is a particular radius at which the differential flux profile is most tightly constrained, and another (slightly larger) angle within which the total annihilation flux is best constrained \citep{walker11,2011MNRAS.418.1526C,2015MNRAS.446.3002B}. The uncertainty in the flux within $0.01\degg$ is about a factor of 5 and decreases to about 20\% when integrating within about $0.3\degg$, an angle corresponding to twice the projected half-light radius. For Segue 1, however, the situation is somewhat different.  While the integrated $J$ value within $0.01\degg$ can be inferred to within a factor of 6, similar to the case of Draco, and the minimum uncertainty again occurs when integrating within about twice the half-light radius ($\theta \approx 0.15\degg$), even there $J$ can only be determined to within a factor of 3.5.  We do not see the drastic decrease in the uncertainty of Segue 1's expected emission that we see with most of the classical dwarfs.  The larger uncertainty for Segue 1 is a direct consequence of the relatively small size of its available kinematic sample.

We can quantify the extent to which halos can be spatially resolved in gamma-ray telescopes by comparing the derived emission profiles for either annihilation or decay with the point spread function (PSF) of specific instruments.

\begin{figure*}
\includegraphics[scale=0.7]{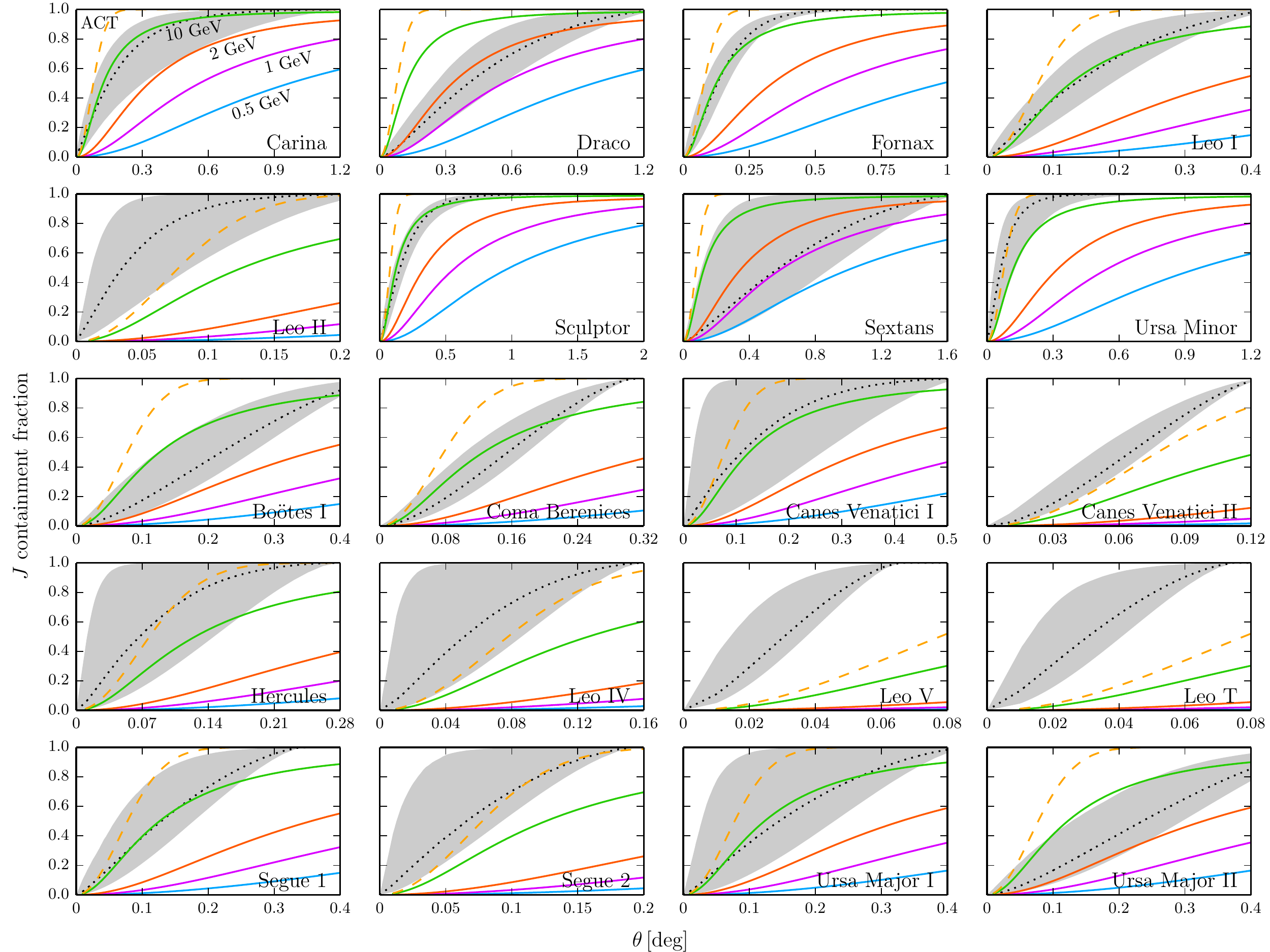}
  \caption{The containment fraction for annihilation as a function of angular distance from the center of each dwarf. The $x$-axis is in degrees up to $\thetamax$ for each dwarf. The containment fraction is defined as $J(\theta) / J(\thetamax)$. The dotted line shows the median value of the containment fraction while the shaded band corresponds to its 16$^{\rm th}$ and 84$^{\rm th}$ percentiles at each angle. The solid blue, magenta, red, and green lines show the containment fraction of the Fermi-LAT PSF at 0.5, 1, 2 \& 10 GeV respectively. The dashed orange line corresponds to the PSF of a typical ACT (68\% containment of $0.1\degg$). This figure together with Fig.~\ref{fig:fig_Jtot_all} can be used to estimate the proper normalization of the expected emission within any aperture for each dwarf. The data used to construct this figure and Fig.~\ref{fig:fig_MvsPSF} are available in machine-readable form as described in Appendix~\ref{sec:appendix}.}
  \label{fig:fig_JvsPSF}
\end{figure*}

\begin{figure*}
\includegraphics[scale=0.7]{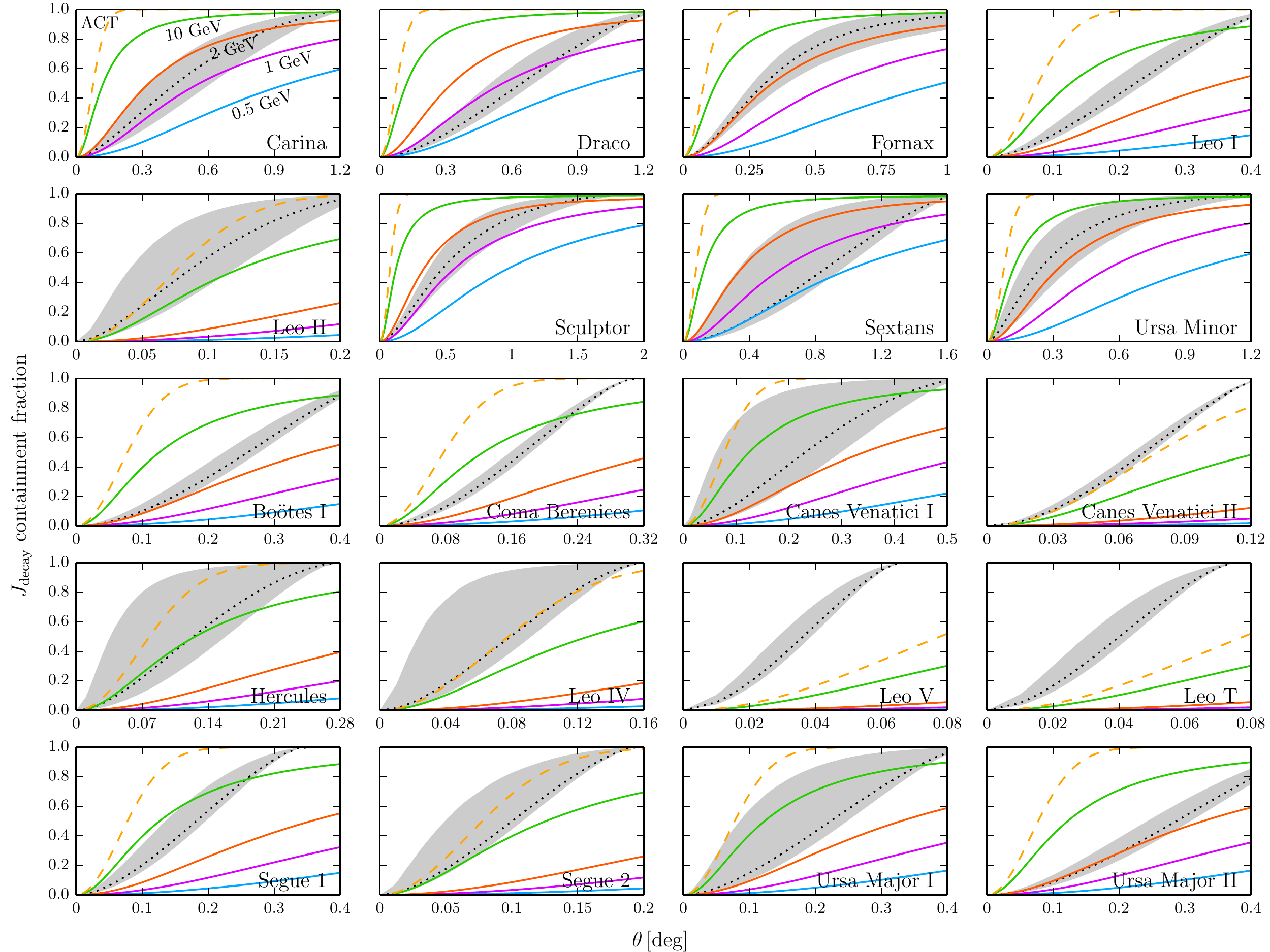}
  \caption{Same as Fig.~\ref{fig:fig_JvsPSF} but for dark matter decay.}
  \label{fig:fig_MvsPSF}
\end{figure*}

Figures~\ref{fig:fig_JvsPSF} and \ref{fig:fig_MvsPSF} show the angular distribution of dark matter annihilation and decay. The bands show constraints on the ``containment fraction'' curves for the different dwarfs. The containment fraction, at angle $\theta$, is defined simply as $J(\theta) / J(\thetamax)$, where $J(\theta)$ is given by Eq.~\eqref{eqn:defJtheta}. Each halo profile gives rise to a containment fraction curve and the dotted line corresponds to the median value of the containment fraction among all the allowed halos, computed at each $\theta$. The shaded band corresponds to the 16$^{\rm th}$ and 84$^{\rm th}$ percentiles. For example, the constraint on the ``half-light radius'' of the dark matter emission profile is the intersection of the horizontal line $y=0.5$ with the shaded band. We use $\theta_{0.5}$ and $\theta_{0.5\, \rm decay}$ to denote the half-light radii for $J$- and $J_{\rm decay}$-profiles and tabulate them in Table~\ref{tab:Jtable}.

The curves in Figs.~\ref{fig:fig_JvsPSF} and \ref{fig:fig_MvsPSF} illustrate the point spread functions (PSFs) of two gamma-ray experiments. The containment fraction of a PSF is simply the probability that a gamma-ray will be reconstructed within an angle $\theta$ of its true origin. The solid blue, magenta, red, and green lines correspond to the PSF of the Fermi-LAT at photon energies of 0.5, 1, 2, and 10 GeV (computed using {\tt gtpsf} --- see software and documentation at the Fermi Science Support Center\footnote{{\tt http://fermi.gsfc.nasa.gov/ssc/data/analysis/}}). The dashed orange line corresponds to a 2-dimensional Gaussian PSF with a 68\% containment angle of $0.1\degg$ (e.g. a Rayleigh distribution with a mean of $0.083\degg$). This corresponds to the benchmark PSF of current-generation Atmospheric \v{C}erenkov Telescopes (ACTs). Figure~\ref{fig:fig_MvsPSF} is identical to Fig.~\ref{fig:fig_JvsPSF} but shows the containment fractions for $\Jdecay$.

We find that for many of the classical dwarfs (Carina, Draco, Fornax, Leo I, Sculptor, Sextans) ACTs should be able to detect extended emission from dark matter annihilation (if the emission can be detected at all) and similarly for some of the ultra-faint dwarfs (Bo\"{o}tes I, Coma Berenices, and Ursa Major II). Regarding Fermi-LAT, at the highest energies ($>10$ GeV) only Draco and, perhaps, Ursa Major II appear to be extended enough to be detected, and therefore any limits derived using Fermi-LAT data will not be affected significantly by the assumption of point sources when it comes to dwarf galaxies (in agreement with \citet{2014PhRvD..89d2001A}).

\vspace{1cm}

\section{Comparison with other work}
\label{sec:others}

\begin{figure*}
  \epsscale{1.2}
  \plotone{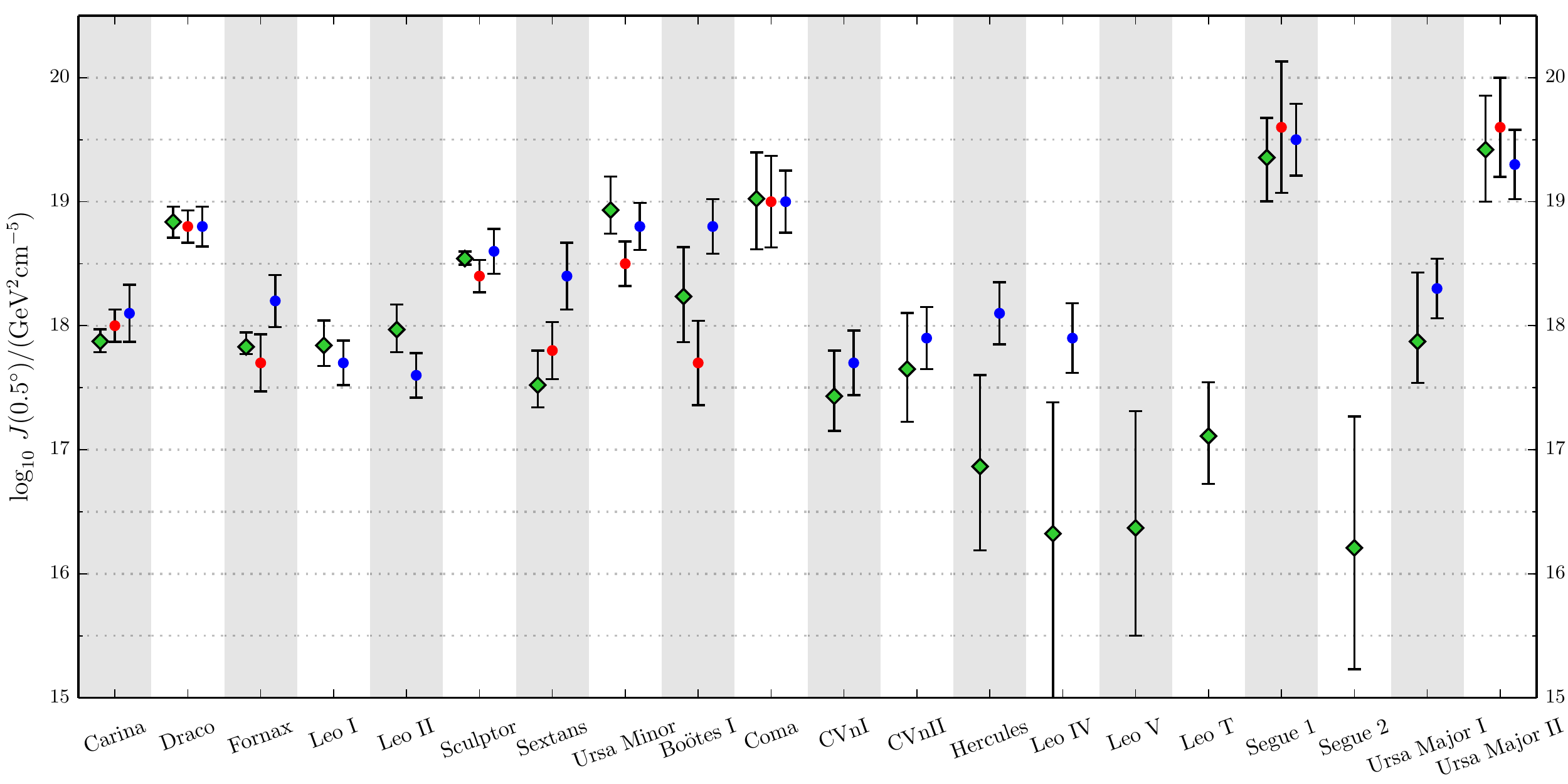}
  \caption{$J$-profiles integrated within $0.5\degg$ for all the dwarf galaxies. The green diamonds with error bars show the range of $J$ values derived in this work with $1\sigma$ error bars. The red and blue points are the integrated $J$ values for NFW profiles reported in~\citet{2011PhRvL.107x1302A} and~\citet{2014PhRvD..89d2001A}, respectively, with error bars corresponding to the $1\sigma$ uncertainties. }
  \label{fig:compareallJ}
\end{figure*}

In order to compare the expected signals derived in this paper with the predictions from other work, Figure~\ref{fig:compareallJ} shows the distributions of the $J$-profile integrated within a cone of radius $0.5\degg$ for all the dwarf galaxies in the sample. In this figure, the green diamonds are the median values of $J$ from the sampled halos in this work with $\pm 1 \sigma$ error bars. The red and blue points show the $J$ values integrated within $0.5\degg$ reported by~\citet{2011PhRvL.107x1302A} and~\citet[][NFW profiles]{2014PhRvD..89d2001A} respectively. The $J$ values in the latter study come from~\citet{2013arXiv1309.2641M}. The error bars on these points correspond to the $1\sigma$ errors quoted in those studies.

We find that to within an order of magnitude the constraints on $J$ values are consistent with those derived by \citet{2011PhRvL.107x1302A} and \citet{2014PhRvD..89d2001A,2013arXiv1309.2641M}. The differences, however, appear to be systematic and not random. In particular, for the ultra-faint dwarfs the central $J$ values from \citet{2014PhRvD..89d2001A,2013arXiv1309.2641M} are almost always larger than those we find. Of the eight classical dwarfs, which are the least dependent on priors in the analysis presented here, the results are inconsistent by more than $1\sigma$ for Fornax, Leo II, and Sextans when compared to~\citet{2014PhRvD..89d2001A}. For Fornax and Ursa Minor the results of \citet{2011PhRvL.107x1302A} also disagree at similar significance.

\citet{2013arXiv1309.2641M} introduced a new analysis that effectively models the entire population of Milky Way dwarfs simultaneously.  As with the study of \citet{2011PhRvL.107x1302A}, all halos are assumed to follow the NFW form, but the (assumed linear) relationship between $\log \vmax$ and $\log \rmax$ is modeled simultaneously with an (assumed linear) relationship between $\log L$ and $\log\vmax$, where $L$ is optical luminosity.  Empirical information about kinematics enters only in the form of published estimates of masses enclosed within dwarf half-light radii, which are determined given $\vmax$ and $\rmax$.  With respect to the $J$ values estimated by \cite{2013arXiv1309.2641M}, the analysis presented here yields tighter constraints for the classical dwarfs and looser constraints for ultra-faints.  The former makes intuitive sense as the analysis presented here uses the greater amount of information that is available in the larger, unbinned kinematic data sets that are available for classical dwarfs (e.g. using velocity dispersion as a function of radius).  We suspect that the discrepency for the ultra-faints results from the assumption by \citet{2013arXiv1309.2641M} that the scatter about the linear relation between $\log L$ and $\log\vmax$ is independent of luminosity and that between $\log\vmax$ and $\log\rmax$ is independent of $\log\vmax$.  Given the hierarchical nature of the model, this assumption enables ``sharing'' of information among classical and ultra-faint dwarfs.  That is, the model is constrained primarily by the better-sampled classical dwarfs, but because scatter about the modeled relations is assumed to be independent of luminosity, it is impossible for the inferred constraints for ultra-faints to be looser than those inferred at the luminous end (notice the uniformity of error bars among nearly all the blue points in Figure \ref{fig:compareallJ}, especially for the ultra-faints).  Thus, it may be possible that the error bars obtained by \citet{2013arXiv1309.2641M} for the ultra-faint dwarfs are suppressed by the unduly restrictive hierarchical model priors.  

\section{Conclusion}
\label{sec:conclusion}

Dwarf galaxies represent the cleanest laboratory in which to search for a signal from dark matter annihilation and decay. Understanding the dark matter distribution in these systems is therefore of paramount importance. 

Using the latest kinematic observations from 20 dwarf galaxies, we have presented a thorough study of the dark matter content and distribution in these systems. 
We have shown that, owing to the quality and size of available data sets, the classical dwarf galaxies are better constrained than ultra-faint dwarfs. Relevant to dark matter annihilation searches, Draco and Ursa Minor are the most promising dwarfs, with Sculptor being only a factor 3 fainter but with the smallest uncertainty in its emission. From the ultra-faint dwarf sample, Segue 1, Ursa Major II, and Coma Berenices are by far the brightest dwarf galaxies, albeit with significantly larger systematic uncertainties.  Relevant to dark matter decay studies, Draco is the brightest dwarf galaxy in the sample, with Sculptor, Sextans and Ursa Major II a factor of roughly 3 fainter.

In addition, we have explored the angular distribution of annihilation and decay emission from these dwarfs, and show that Atmospheric \v{C}erenkov Telescopes have the necessary angular resolution to detect extended emission from many of the classical and ultra-faint dwarf galaxies. With the future of gamma-ray astronomy efforts focused on the construction of the Cherenkov Telescope Array (CTA) \citep{2011ExA....32..193A}, this work provides the necessary ingredients that can be used to derive the expected reach of CTA in the context of dark matter searches.

In summary, the results presented here provide one of the key pieces of information required to test hypotheses about dark matter annihilation and decay using dwarf galaxies. Such searches are currently even more important in light of the recent claims of dark matter annihilation evidence from the Galactic center region. Dwarf galaxies are the most reliable sources where these claims can be tested. This shall be the focus of Paper II.

\acknowledgements
We acknowledge useful conversations with Gordon Blackadder, Vincent Bonnivard, Celine Combet, David Maurin, Justin Read, Louie Strigari, Mark Wilkinson, and Andrew Zentner. SMK is supported by DOE DE-SC0010010, NSF PHYS-1417505 and NASA NNX13AO94G.  MGW acknowledges support from NSF grant AST-1313045.  SMK and MGW thank the Aspen Center for Physics for hospitality where part of this work was completed. 

\appendix
\section{Tabulated constraints for individual dwarfs}
\label{sec:appendix}
In this appendix we present an additional table with integrated $J$-profiles and containment fractions for each dwarf as a function of $\theta$. The data in this tables can be used to reconstruct Figs.~\ref{fig:fig_Jtot_all}, \ref{fig:fig_Jdecaytot_all}, \ref{fig:fig_JvsPSF}, \ref{fig:fig_MvsPSF}, \ref{fig:compareallJ}, the lower panels of Fig.~\ref{fig:fig_profiles_Draco_Seg1}, and Table~\ref{tab:Jtable}.

The first few lines of the table are shown in Table~\ref{tab:dataCarina}. The full table, containing all the dwarfs, is available in machine-readable format as an ancillary file. For each value of $\theta$ there are four constrained quantities: the integrated $J$-profiles $J(\theta)$ and $J_{\rm decay}(\theta$), and the containment fractions $J(\theta)/J(\thetamax)$ and $J_{\rm decay}(\theta)/J_{\rm decay}(\thetamax)$. The columns labeled $-2\sigma, -1\sigma, \dots, +2\sigma$ correspond to quantiles of the distribution of a quantity among the Monte Carlo realizations of halos. For example, for each halo realization we find the value of $J(0.5\degg)$. The $2.3, 16, 50, 84$, and $97.7$ percentiles of this collection of $J(0.5\degg)$'s are listed in columns 3--7 of the row corresponding to $\theta=0.5\degg$. The table has 50 rows per dwarf, corresponding to log-spaced angles between $0.01\degg$ and $2.6\degg$ (the largest $\thetamax$ of any of the dwarfs).

\section{Halo profile parameter constraints}

In order to facilitate the reproducibility of these results we also provide constraints on the individual halo profile parameters for each dwarf. For each of the parameters $\rho_s, r_s, \alpha, \beta, \gamma,$ and $\beta_a$ (see Eqs.~\ref{eq:rho} and~\ref{eq:beta_a}), we provide the $-2\sigma, -1\sigma$, median, $+1\sigma$, and $+2\sigma$ quantiles of the posterior distribution (marginalized over the other parameters). The consistency conditions described in Secs.~\ref{sec:tidalradius} and~\ref{sec:cosmo} have been applied to the posterior samples. Table~\ref{tab:profileparams} shows the constraints for the first few dwarfs. The full table is available in machine-readable format as an ancillary file.

\clearpage
\begin{landscape}

\setlength{\tabcolsep}{0.03in}
\begin{deluxetable*}{c | c | ccccc | ccccc | ccccc | ccccc}
  \tabletypesize{\scriptsize} 
  \tablewidth{0pc}
  \tablecolumns{21}
  \tablecaption{\scriptsize Integrated J-profiles and Containment Fractions\tablenotemark{*}}
  \tablehead{
  \colhead{Name} &
  \colhead{$\theta$} & \multicolumn{5}{c}{$\log_{10} J(\theta)$} & \multicolumn{5}{c}{$\log_{10} J_{\rm decay}(\theta)$} & \multicolumn{5}{c}{$J(\theta)/J(\thetamax)$} & \multicolumn{5}{c}{$J_{\rm decay}(\theta)/J_{\rm decay}(\thetamax)$} \\
\colhead{} &  \colhead{$[\deg]$} & \multicolumn{5}{c}{$\mathrm{[GeV^2 \,cm^{-5}]}$} & \multicolumn{5}{c}{$\mathrm{[GeV \,cm^{-2}]}$} & \multicolumn{5}{c}{---} & \multicolumn{5}{c}{---} \\
 \colhead{}& \colhead{} & $-2\sigma$ & $-1\sigma$ & median & $+1\sigma$ & $+2\sigma$ & $-2\sigma$ & $-1\sigma$ & median & $+1\sigma$ & $+2\sigma$ & $-2\sigma$ & $-1\sigma$ & median & $+1\sigma$ & $+2\sigma$ & $-2\sigma$ & $-1\sigma$ & median & $+1\sigma$ & $+2\sigma$
  }
  \startdata
Carina & 0.010 & 15.38 & 15.88 & 16.52 & 16.93 & 17.33 & 14.98 & 15.09 & 15.23 & 15.32 & 15.44 & 2.3e-03 & 8.8e-03 & 3.8e-02 & 1.0e-01 & 2.3e-01 & 3.2e-04 & 5.4e-04 & 1.1e-03 & 2.3e-03 & 4.6e-03 \\
Carina & 0.011 & 15.47 & 15.96 & 16.57 & 16.97 & 17.37 & 15.08 & 15.19 & 15.32 & 15.41 & 15.52 & 2.8e-03 & 1.0e-02 & 4.2e-02 & 1.1e-01 & 2.5e-01 & 4.0e-04 & 6.7e-04 & 1.3e-03 & 2.8e-03 & 5.6e-03 \\
Carina & 0.013 & 15.57 & 16.03 & 16.62 & 17.01 & 17.40 & 15.17 & 15.28 & 15.41 & 15.49 & 15.60 & 3.5e-03 & 1.2e-02 & 4.7e-02 & 1.3e-01 & 2.7e-01 & 5.0e-04 & 8.2e-04 & 1.6e-03 & 3.4e-03 & 6.7e-03 \\
Carina & 0.014 & 15.66 & 16.11 & 16.67 & 17.04 & 17.44 & 15.27 & 15.38 & 15.49 & 15.57 & 15.68 & 4.3e-03 & 1.5e-02 & 5.3e-02 & 1.4e-01 & 2.9e-01 & 6.2e-04 & 1.0e-03 & 2.0e-03 & 4.2e-03 & 8.1e-03 \\
Carina & 0.016 & 15.75 & 16.19 & 16.72 & 17.08 & 17.47 & 15.37 & 15.47 & 15.58 & 15.66 & 15.76 & 5.3e-03 & 1.7e-02 & 6.0e-02 & 1.5e-01 & 3.1e-01 & 7.6e-04 & 1.2e-03 & 2.5e-03 & 5.1e-03 & 9.8e-03 \\
Carina & 0.018 & 15.84 & 16.26 & 16.77 & 17.11 & 17.50 & 15.46 & 15.56 & 15.67 & 15.74 & 15.84 & 6.4e-03 & 2.1e-02 & 6.7e-02 & 1.6e-01 & 3.3e-01 & 9.5e-04 & 1.5e-03 & 3.0e-03 & 6.2e-03 & 1.2e-02 \\
Carina & 0.020 & 15.93 & 16.34 & 16.82 & 17.15 & 17.53 & 15.56 & 15.66 & 15.76 & 15.83 & 15.92 & 7.7e-03 & 2.4e-02 & 7.5e-02 & 1.8e-01 & 3.6e-01 & 1.2e-03 & 1.9e-03 & 3.7e-03 & 7.5e-03 & 1.4e-02 \\
\dots
  \enddata
  \tablenotetext{*}{This table is published in its entirety in the electronic edition of the Astrophysical Journal. A portion is shown here for guidance regarding its form and content.}
  \label{tab:dataCarina}
\end{deluxetable*}

\setlength{\tabcolsep}{0.01in}
\begin{deluxetable*}{c | ccccc | ccccc | ccccc | ccccc | ccccc | ccccc}
  \tabletypesize{\scriptsize} 
  \tablewidth{0pc}
  \tablecolumns{21}
  \tablecaption{\scriptsize Halo Profile Parameter Constraints\tablenotemark{*}}
  \tablehead{
  \colhead{Name} & \multicolumn{5}{c}{$\log_{10} \rho_s$} & \multicolumn{5}{c}{$\log_{10} r_s$} & \multicolumn{5}{c}{$\alpha$} & \multicolumn{5}{c}{$\beta$} & \multicolumn{5}{c}{$\gamma$} & \multicolumn{5}{c}{$-\log_{10}(1-\beta_a)$} \\
  \colhead{---} & \multicolumn{5}{c}{$\mathrm{[M_\odot pc^{-3}]}$} & \multicolumn{5}{c}{$\mathrm{[pc]}$} & \multicolumn{5}{c}{---} & \multicolumn{5}{c}{---} & \multicolumn{5}{c}{---} & \multicolumn{5}{c}{---} \\
 \colhead{} & $-2\sigma$ & $-1\sigma$ & median & $+1\sigma$ & $+2\sigma$ & $-2\sigma$ & $-1\sigma$ & median & $+1\sigma$ & $+2\sigma$ & $-2\sigma$ & $-1\sigma$ & median & $+1\sigma$ & $+2\sigma$ & $-2\sigma$ & $-1\sigma$ & median & $+1\sigma$ & $+2\sigma$ & $-2\sigma$ & $-1\sigma$ & median & $+1\sigma$ & $+2\sigma$ & $-2\sigma$ & $-1\sigma$ & median & $+1\sigma$ & $+2\sigma$
  }
  \startdata

Carina & -3.74 & -2.86 & -1.96 & -1.28 & -0.73 &  2.61 &  2.90 &  3.31 &  3.96 &  4.63 &  0.62 &  0.87 &  1.47 &  2.41 &  2.89 &  3.12 &  3.75 &  5.60 &  8.36 &  9.71 &  0.13 &  0.54 &  0.95 &  1.13 &  1.19 & -0.41 & -0.23 & -0.07 &  0.08 &  0.25 \\ 
Draco & -3.36 & -2.66 & -1.74 & -1.09 & -0.78 &  2.85 &  3.11 &  3.57 &  4.34 &  4.83 &  0.75 &  1.18 &  2.01 &  2.65 &  2.95 &  3.19 &  4.16 &  6.34 &  8.69 &  9.74 &  0.06 &  0.29 &  0.71 &  1.02 &  1.16 &  0.01 &  0.25 &  0.54 &  0.81 &  0.97 \\ 
Fornax & -2.11 & -1.83 & -1.49 & -1.20 & -0.76 &  2.73 &  2.93 &  3.09 &  3.25 &  3.51 &  0.86 &  1.39 &  2.13 &  2.73 &  2.96 &  3.30 &  4.54 &  6.97 &  9.02 &  9.84 &  0.03 &  0.19 &  0.61 &  1.02 &  1.17 & -0.25 & -0.15 & -0.06 &  0.02 &  0.09 \\ 
Leo I & -3.69 & -3.08 & -2.18 & -1.38 & -0.92 &  2.91 &  3.23 &  3.80 &  4.55 &  4.91 &  0.71 &  1.12 &  1.93 &  2.64 &  2.94 &  3.16 &  3.99 &  6.15 &  8.70 &  9.76 &  0.09 &  0.41 &  0.84 &  1.08 &  1.18 & -0.20 & -0.03 &  0.14 &  0.37 &  0.72 \\
Leo II & -3.24 & -2.31 & -0.92 & -0.03 &  0.47 &  1.95 &  2.29 &  2.89 &  4.03 &  4.77 &  0.64 &  1.01 &  1.76 &  2.53 &  2.90 &  3.16 &  3.89 &  5.95 &  8.56 &  9.73 &  0.08 &  0.35 &  0.82 &  1.08 &  1.18 & -0.88 & -0.51 & -0.01 &  0.51 &  0.88 \\
\dots
  \enddata
  \tablenotetext{*}{This table is published in its entirety in the electronic edition of the Astrophysical Journal. A portion is shown here for guidance regarding its form and content.}
  \label{tab:profileparams}
\end{deluxetable*}

\clearpage
\end{landscape}

\bibliography{manuscript}

\end{document}